\documentclass[11pt]{article}
\title{Exact Posterior Mean and Covariance for Generalized Linear Mixed Models}
\author{Tonglin Zhang\footnote{Department of Statistics, Purdue University, 150 North University Street, West Lafayette, IN 47907-2066, Email: tlzhang@purdue.edu} \\Purdue University}
\usepackage{epsfig}
\usepackage{mathrsfs}
\usepackage{amsfonts}
\usepackage{amsmath}
\usepackage{wrapfig,lipsum,booktabs}
\usepackage{bm}
\usepackage{algorithm}
\usepackage{algpseudocode}
\def\qed{\hfill$\diamondsuit$}

\newtheorem{thm}{Theorem}
\newtheorem{cor}{Corollary}
\newtheorem{lem}{Lemma}
\begin{document}
\maketitle
\def\eqalign#1{\null\,\vcenter{\openup\jot\ialign
              {\strut\hfil$\displaystyle{##}$&$\displaystyle{{}##}$
               \hfil\crcr#1\crcr}}\,}

\setcounter{page}{1}

\begin{abstract}
A novel method is proposed for the exact posterior mean and covariance of the random effects given the response in a generalized linear mixed model (GLMM) when the response does not follow normal. The research solves a long-standing problem in Bayesian statistics when an intractable integral appears in the posterior distribution. It is well-known that the posterior distribution of the random effects given the response in a GLMM when the response does not follow normal contains intractable integrals. Previous methods rely on Monte Carlo simulations for the posterior distributions. They do not provide the exact posterior mean and covariance of the random effects given the response. The special integral computation (SIC) method is proposed to overcome the difficulty. The SIC method does not use the posterior distribution in the computation. It devises an optimization problem to reach the task. An advantage is that the computation of the posterior distribution is unnecessary. The proposed SIC avoids the main difficulty in Bayesian analysis when intractable integrals appear in the posterior distribution. 
\end{abstract}



\section{Introduction}
\label{sec:introduction}

A generalized linear mixed model (GLMM) is a special case of Bayesian hierarchical models (BHMs) with the first level specified for the conditional distribution of the response given the random effects and the second level specified for the prior distribution of the random effects.  Except for a response following a multivariate normal distribution given the random effects, all the remaining cases of GLMMs contain intractable integrals in their marginal distributions, leading to difficulty in the Bayesian analysis of GLMMs. We investigate the posterior mean and covariance problem. We find that the exact posterior mean and covariance of the random effects given the response can be calculated by optimization. The computation does not involve an explicit expression or numerical approximation of the posterior distribution. A nice property is that we do not use any sampling (e.g., MCMC) methods, implying that our method is not simulation-based. We solve the posterior mean and covariance problem, although the posterior distribution problem remains unsolved. 

A GLMM is formulated by incorporating normally distributed random effects into linear components of a generalized linear model (GLM) with a response ${\bm y}=(y_1,\dots,y_n)^\top\in\mathbb{R}^n$ following an exponential family distribution. Suppose that the dispersion parameter is not involved such that the probability density function (PDF) or probability mass function (PMF) of ${\bm y}$ can be expressed as
\begin{equation}
\label{eq:exponential family distribution}
f({\bm y}|{\bm\gamma})=\exp[{\bm y}^\top{\bm\eta}-{\bf 1}^\top b({\bm\eta})+{{\bf 1}^\top}c({\bm y})],
\end{equation}
where ${\bm\eta}=(\eta_1,\dots,\eta_n)^\top$ is an $n$-dimensional vector, $b({\bm\eta})=(b(\eta_1),\dots,b(\eta_n))^\top$ is derived by a real transformation $b(\cdot)$ on ${\bm\eta}$,  and $c({\bm y})=(c(y_1),\dots,c(y_n))^\top$ is a normalized constant vector. The mathematical formulation given by~\eqref{eq:exponential family distribution} is appropriate if ${\bm y}$ follows a binomial, Poisson, or multinomial distribution. A GLMM with the canonical link is
\begin{equation}
\label{eq:generalized linear mixed model}
{\bm\eta}={\bf X}{\bm\beta}+{\bf Z}{\bm\gamma}
\end{equation}
and 
\begin{equation}
\label{eq:distribution of random effects}
{\bm\gamma}\sim{\cal N}({\bf 0},{\bf D}),
\end{equation}
where ${\bf D}={\bf D}_{\bm\omega}$ is determined by a hyperparameter vector ${\bm\omega}$ for the variance components, ${\bf X}=({\bm x}_1^\top,\dots,{\bm x}_n^\top)^\top\in\mathbb{R}^{n\times p}$ with ${\bm x}_{i}=(x_{i0},\dots,x_{i(p-1)})^\top\in\mathbb{R}^p$ is a design matrix for fixed effects, ${\bm\beta}=(\beta_0,\dots,\beta_{p-1})^\top\in\mathbb{R}^p$ is a parameter vector for fixed effects, ${\bf Z}=({\bm z}_1^\top,\dots,{\bm z}_n^\top)^\top\in\mathbb{R}^{n\times r}$ with ${\bm z}_{i}=(z_{i1},\dots,z_{ir})^\top\in\mathbb{R}^r$ is a design matrix for random effects, and ${\bm\gamma}=(\gamma_1,\dots,\gamma_r)^\top\in\mathbb{R}^{r}$ is a random effects vector. Both ${\bf X}$ and ${\bf Z}$ are full rank. They satisfy ${\rm rank}({\bf X})=p$ and ${\rm rank}({\bf Z})=r$ in~\eqref{eq:generalized linear mixed model}. To specify the GLMM, a prior distribution for ${\bm\gamma}$ is needed. It is assumed that ${\bm\gamma}$ follows a multivariate normal distribution, such that the prior density $\pi({\bm\gamma})$ for ${\bm\gamma}$ can be expressed as the PDF of ${\mathcal N}({\bf 0},{\bf D})$ given by~\eqref{eq:distribution of random effects}. In practice, ${\bf D}$ is often specified by longitudinal data or spatial models. 

The exponential family distribution given by~\eqref{eq:exponential family distribution} satisfies ${\bm\mu}=(\mu_1,\dots,\mu_n)^\top=b'({\bm\eta})={\rm E}({\bm y}|{\bm\eta})$ and ${\rm cov}({\bm y}|{\bm\eta})={\rm diag}\{b''({\bm\eta})\}$, where $\mu_i=b'(\eta_i)={\rm E}(y_i|{\bm\eta})$ and ${\rm var}(y_i|{\bm\eta})=b''(\eta_i)$. Under~\eqref{eq:generalized linear mixed model} and~\eqref{eq:distribution of random effects}, the GLMM can be expressed as ${\bm y}|{\bm\gamma}\sim f({\bm y}|{\bm\gamma})$ and ${\bm\gamma}\sim\pi({\bm\gamma})$. 

The GLMM jointly defined by~\eqref{eq:exponential family distribution},~\eqref{eq:generalized linear mixed model}, and~\eqref{eq:distribution of random effects} is a special case of BHMs with two hierarchical levels. The first level, specified by~\eqref{eq:exponential family distribution} and \eqref{eq:generalized linear mixed model}, provides $f({\bm y}|{\bm\gamma})$ the conditional distribution of ${\bm y}|{\bm\gamma}$. The second level, specified by \eqref{eq:distribution of random effects}, provides ${\bm\gamma}\sim\pi({\bm\gamma})$ the prior distribution for the random effects. The choice of the multivariate normal distribution for ${\bm\gamma}$ is convenient for modeling dependencies between the random effects. 

The posterior density of ${\bm\gamma}$ is 
\begin{equation}
\label{eq:posterior distribution of gamma given y}
q({\bm\gamma}|{\bm y})={f({\bm y}|{\bm\gamma}) \pi({\bm\gamma}) \over \bar f({\bm y})},
\end{equation}
where 
\begin{equation}
\label{eq:likelihood function of the response y}
\bar f({\bm y})=\int_{\mathbb{R}^r}f({\bm y}|{\bm\gamma})\pi({\bm\gamma})d{\bm\gamma}
\end{equation}
is the marginal PDF or PMF  of ${\bm y}$ (i.e., the likelihood function). The posterior mean of ${\bm\gamma}$ is $r$-dimensional vector as 
\begin{equation}
\label{eq:definition of posterior mean vector of random effects}
{\bm\xi}={\bm\xi}_{\bm y}={\rm E}({\bm\gamma}|{\bm y})=\int_{\mathbb{R}^r} {\bm\gamma}q({\bm\gamma}|{\bm y})d{\bm\gamma}.
\end{equation}
The posterior covariance of ${\bm\gamma}$ is $r\times r$ matrix as
\begin{equation}
\label{eq:definition of posterior covariance of random effects}
{\bm\Xi}={\bm\Xi}_{\bm y}={\rm E}[({\bm\gamma}-{\bm\xi})^\top({\bm\gamma}-{\bm\xi})]=\int_{\mathbb{R}^r} ({\bm\gamma}-{\bm\xi})^\top({\bm\gamma}-{\bm\xi})q({\bm\gamma}|{\bm y})d{\bm\gamma}.
\end{equation}

We solve the posterior mean problem given by~\eqref{eq:definition of posterior mean vector of random effects} and the posterior covariance problem given by~\eqref{eq:definition of posterior covariance of random effects} but not the posterior distribution problem given by~\eqref{eq:posterior distribution of gamma given y}. The marginal distribution problem given by~\eqref{eq:likelihood function of the response y} remains unsolved. 

It is well-known that the integral for $\bar f({\bm y})$ given by~\eqref{eq:likelihood function of the response y} is intractable if $f({\bm y}|{\bm\gamma})$ given by~\eqref{eq:exponential family distribution} is not multivariate normal. The posterior distribution $q({\bm\gamma}|{\bm y})$ given by~\eqref{eq:posterior distribution of gamma given y} cannot be analytically solved. 
It is impossible to directly use~\eqref{eq:definition of posterior mean vector of random effects} and~\eqref{eq:definition of posterior covariance of random effects} to compute the exact values of the posterior mean and covariance of ${\bm\gamma}$ given ${\bm y}$. Previous methods approximately compute ${\bm\xi}$ and ${\bm\Xi}$ by Monte Carlo methods (e.g., MCMC). They do not provide the exact ${\bm\xi}$ and ${\bm\Xi}$. We devise a special integral computation (SIC) method to overcome the difficulty. The SIC designs an optimization problem that can lead to the exact values of ${\bm\xi}$ and ${\bm\Xi}$. Because neither exact nor approximate $q({\bm\theta}|{\bm y})$ is needed, the SIC does not suffer from the computational difficulty caused by the intractable integral on the right-hand side of~\eqref{eq:likelihood function of the response y}. The details of our method are presented in Section~\ref{sec:main result}.

The intractable integral problem in Bayesian analysis is connected with Louville's Theorem in differential algebra, a field of mathematics. The corresponding mathematical problem is called integration in closed forms, meaning that an integration can be expressed as elementary functions. French mathematician Joseph Liouville first studied this problem in 1833~\cite{liouville1933a,liouville1933b,liouville1933c}, about antiderivatives to be elementary functions. The impact of Liouville's work received little attention until the 1940s when Ritt~\cite{ritt1948} developed a field called {\it differential algebra} in mathematics. A growing interest following Liouville's work has appeared with the advances of computer languages of symbolic mathematical computation. The mathematics of the indefinite integration problem has evolved significantly since the work of~\cite{risch1969,rosenlicht1972}. The domain of the integrals on the right-hand sides of~\eqref{eq:likelihood function of the response y},~\eqref{eq:definition of posterior mean vector of random effects}, and~\eqref{eq:definition of posterior covariance of random effects} are the entire $\mathbb{R}^r$. They should not be classified as indefinite integrals in mathematics. We treat them as special integrals. The special integral techniques can be used. We derive our method based on these techniques. 

The article is organized as follows. In Section~\ref{sec:liouville's theorem in mathematics}, we review Liouville's Theorem in mathematics. In Section~\ref{sec:main result}, we present our main result. In Section~\ref{sec:prediction}, we propose a prediction method for the random effects. In Section~\ref{sec:example}, we provide a few examples. In Section~\ref{sec:simulation}, we evaluate the performance of our method by Monte Carlo simulations. In Section~\ref{sec:application}, we apply our method to a real-world dataset. In Section~\ref{sec:conclusion}, we conclude the article. We put all proofs in the Appendix.

\section{Liouville's Theorem in Mathematics}
\label{sec:liouville's theorem in mathematics}

Liouville's Theorem answers a question in elementary calculus that the antiderivative of an elementary function may not be expressed as a certain elementary function. This is called the nonelementary antiderivative problem in differential algebra, a field of mathematics that appeared after the work of~\cite{ritt1948}. In mathematics, an elementary function is a function of a single variable that can be expressed as sums, products, roots, and compositions of constants, rational powers, exponential, logarithms, trigonometric, inverse trigonometric, hyperbolic, inverse hyperbolic functions as well as their finite additions, subtractions, multiplications, divisions, roots, and compositions~\cite[Chapter 12, e.g.]{geddes1992}. 

When solving for an indefinite integral of a given elementary function $f(x)$, the fundamental problem of elementary calculus is to answer whether the antiderivative $F(x)=\int_a^x f(t)dt$ can be expressed as an elementary function. This is the problem of integration in closed form or the problem of integration in finite terms, well-known in differential algebra. The ideal case is that there is an elementary function identical to $F(x)$ for any $x$ in its domain. Liouville's Theorem states that the ideal case may not be achieved due to the non-existence of an elementary function for $F(x)$. 

Given that $F(x)$ is nonelementary, $F(x)$ may still be solved in special cases. A typical example is $f(x)=x^{-1}\sin{x}$. It has been shown that $F(x)=\int_0^x t^{-1}\sin{t}dt$ as an indefinite integral is nonelementary~\cite{williams1993}. As a special integral, $F(\infty)=\int_0^\infty t^{-1}\sin{t}dt$ can be solved via $F(\infty)=\lim_{\alpha\rightarrow 0^+} \int_0^\infty e^{-\alpha{t}}t^{-1}\sin{t}dt$. Let $h(\alpha)=\int_0^\infty e^{-\alpha{t}} t^{-1}\sin{t} dt$. By $h'(\alpha)=-\int_0^\infty e^{-\alpha{t}} \sin{t} dt=-1/(1+\alpha^2)$ and $h(\infty)=0$, there is $h(\alpha)= \int_\alpha^\infty h'(t)dt={\pi/ 2}-\arctan\alpha$. The definite integral is
\begin{equation}
\label{eq:sin(x)/x special integral}
F(\infty)=\int_0^\infty {\sin{t}\over t}dt={\pi\over 2}.
\end{equation}
The indefinite integral 
\begin{equation}
\label{eq:sin(x)/x antiderivatives}
F(x)=\int_0^x {\sin(t)\over t}dt 
\end{equation}
remains unsolved.

In mathematics, the definite integral given by~\eqref{eq:sin(x)/x special integral} is classified as a special integral (another term is a specific definite integral). The indefinite integral given by~\eqref{eq:sin(x)/x antiderivatives} is classified as a nonelementary antiderivative, also called an intractable integral. The lower or upper bounds of the intractable integrals can be arbitrary. Special integrals are obtained when the lower and upper bounds of the intractable integrals are $0$ or $\pm\infty$. Given that an antiderivative is intractable, the corresponding special integral can only be solved specifically (if it can be solved). Therefore, the method for~\eqref{eq:sin(x)/x special integral} may not work if another special integral is considered.  


\section{Main Result}
\label{sec:main result}

The right-hand sides of~\eqref{eq:likelihood function of the response y},~\eqref{eq:definition of posterior mean vector of random effects}, and~\eqref{eq:definition of posterior covariance of random effects} can be treated as intractable or special integrals. The intractable integral approach attempts to solve the corresponding indefinite integrals with the domain of the integrals to be arbitrary. The right-hand sides of~\eqref{eq:likelihood function of the response y},~\eqref{eq:definition of posterior mean vector of random effects}, and~\eqref{eq:definition of posterior covariance of random effects} are derived if the domain of the integrals grows to be the entire $\mathbb{R}^r$. In mathematics, an intractable integral is often evaluated by a Taylor series. However, the coefficients can still be nonelementary even if the intractable integral has a convergent Taylor series, leading to difficulty in computing the corresponding Taylor series. The special integral approach fixes the domain of an integral to be the entire $\mathbb{R}^r$. It attempts to devise a mathematical problem that can induce the solution of the special integral.  We adopt the special integral approach in our research. We point out that we can work out~\eqref{eq:definition of posterior mean vector of random effects} and~\eqref{eq:definition of posterior covariance of random effects} but not~\eqref{eq:likelihood function of the response y}. 

Because $c({\bm y})$ does not depend on ${\bm\eta}$ in~\eqref{eq:exponential family distribution}, $f_{\bm\alpha}({\bm y}|{\bm\gamma})=\exp[{\bm y}^\top ({\bm\eta}+{\bm\alpha})-{{\bf 1}^\top}b({\bm\eta}+{\bm\alpha})+{{\bf 1}^\top}c({\bm y})]$ satisfying $f_{\bm 0}({\bm y}|{\bm\gamma})=f({\bm y}|{\bm\gamma})$ is a well-defined PDF or PMF of an exponential family distribution for any  ${\bm\alpha}=(\alpha_1,\dots,\alpha_n)\in\mathbb{R}^n$. Let
\begin{equation}
\label{eq:KL divergence for the GLMM}
\eqalign{
h({\bm\alpha})=&{\rm E}\{\log[f_{\bm\alpha}({\bm y}|{\bm\gamma})\pi({\bm\gamma})]\}\cr
=&\int_{\mathbb{R}^r}\int_{\mathbb{R}^n} \log[f_{\bm\alpha}({\bm y}|{\bm\gamma})\pi({\bm\gamma})]f({\bm y}|{\bm\gamma})\pi({\bm\gamma})\nu(d{\bm y})d{\bm\gamma},
}\end{equation}
where $\nu(d{\bm y})$ is the count measure if ${\bm y}$ is discrete or the Lebesgue measure if ${\bm y}$ is continuous. 

The support of $f_{\bm\alpha}({\bm y}|{\bm\gamma})$ does not depend on ${\bm\gamma}$. For an exponential family distribution under a normal prior, all of the conditions of the Lebesgue Dominated Theorem are satisfied. By the Lebesgue Dominated Theorem, we conclude that $h({\bm\alpha})$ is differentiable, leading to  the $i$th component of the gradient vector $\dot{h}({\bm\alpha})$ of $h({\bm\alpha})$ as
\begin{equation}
\label{eq:first-order partial derivative of the KL}
\eqalign{
{\partial{h}({\bm\alpha})\over\partial\alpha_i}=&{\rm E} [y_i-b'(\eta_i+\alpha_i)]\cr
=& \int_{\mathbb{R}^r}\int_{\mathbb{R}^n} [y_i-b'(\eta_i+\alpha_i)]f({\bm y}|{\bm\gamma})\pi({\bm\gamma})\nu(d{\bm y})d{\bm\gamma}.
}\end{equation}
For the Hessian matrix $\ddot h({\bm\alpha})$ of $h({\bm\alpha})$, there is
\begin{equation}
\label{eq:second-order partial derivative of the KL}
\eqalign{
{\partial^2{h}({\bm\alpha})\over\partial\alpha_i^2}=&-{\rm E} [b''(\eta_i+\alpha_i)]\cr
=&- \int_{\mathbb{R}^r}\int_{\mathbb{R}^n} b''(\eta_i+\alpha_i)f({\bm y}|{\bm\gamma})\pi({\bm\gamma})\nu(d{\bm y})d{\bm\gamma}\cr 
}
\end{equation}
and ${\partial^2h({\bm\alpha})/\partial\alpha_i\partial\alpha_j}=0$ for any distinct $i,j\in\{1,\dots,n\}$.

Because $b''(\eta_i+\alpha_i)>0$ for all $i\in\{1,\dots,p\}$, $\ddot{h}({\bm\alpha})$ is negative definite. Note that $\dot h({\bm 0})={\bm 0}$ by ${\rm E}(y_i|{\bm\gamma})=b'(\eta_i)$. The local optimizer of $h({\bm\alpha})$ is unique. The unique maximizer of $h({\bm\alpha})$ is ${\bm\alpha}={\bm 0}$. It is the unique solution of $\dot{h}({\bm\alpha})={\bm 0}$. We obtain
\begin{equation}
\label{eq:global maximizer of h(alpha)}
h({\bm 0})> h({\bm\alpha})
\end{equation} 
for any ${\bm\alpha}\not={\bm 0}$. The choice of ${\bm\alpha}\rightarrow 0$ is flexible in~\eqref{eq:global maximizer of h(alpha)} because $f_{\bm\alpha}({\bm y}|{\bm\gamma})$ is differentiable with respect to ${\bm\alpha}$. We can choose ${\bm\alpha}$ as an arbitrary continuous function of ${\bm y}$, ${\bf X}$, and ${\bm\gamma}$ to reach~\eqref{eq:global maximizer of h(alpha)}. This is an important property in the proof of our main theorem. 

We construct a working linear mixed model (LMM), such that it can induce the same formulations of $\dot h({\bm 0})$ and $\ddot h({\bm 0})$. Observing~\eqref{eq:first-order partial derivative of the KL} and~\eqref{eq:second-order partial derivative of the KL}, we construct the working LMM as
\begin{equation}
\label{eq:working model for posterior mean and covariance of GLMM}
{\bm u}={\bm\alpha}+{\bf X}{\bm\beta}+{\bf Z}{\bm\gamma}+{\bm\epsilon},
\end{equation}
where the prior of ${\bm\gamma}$ is $\pi_{\bm\delta}({\gamma})=\varphi({\bm\gamma};{\bm\delta},{\bf D})$ for a mean vector as ${\bm\delta}=(\delta_1,\dots,\delta_r)^\top\in\mathbb{R}^r$ with $\pi_{\bm 0}({\bm\gamma})=\pi({\bm\gamma})$ given by~\eqref{eq:distribution of random effects}, ${\bm u}={\bm u}_{\bm\eta}=(u_{1},\cdots,u_{n})^\top$ with $u_{i}=u_{i{\bm\eta}}=\eta_i+[y_i-b'(\eta_i)]/b''(\eta_i)$ is the working response vector, ${\bm\epsilon}=(\epsilon_1,\dots,\epsilon_n)^\top$ with $\epsilon_i\sim^{ind}{\cal N}\{0,1/[b''(\eta_i)]\}$ is the working error vector, 
\begin{equation}
\label{eq:normal density}
\varphi({\bm t};{\bm a},{\bf A})={1\over (2\pi)^{d/2}|\det({\bf A})|^{1/2}}e^{-{1\over 2}({\bm t}-{\bm a})^\top{\bf A}^{-1}({\bm t}-{\bm a})}
\end{equation}
 is the PDF of ${\cal N}({\bm a},{\bf A})$, ${\bm a}$ is a $d$-dimensional mean vector, and ${\bf A}$ is a $d\times d$ covariance matrix. Based on~\eqref{eq:working model for posterior mean and covariance of GLMM}, we obtain the following Lamma.

\begin{lem}
\label{lem:normal density posterior identity general}
$ \varphi({\bm u};{\bm\alpha}+{\bf X}{\bm\beta}+{\bf Z}{\bm\gamma},{\bf W}^{-1})\varphi({\bm\gamma};{\bm\delta},{\bf D})=\varphi({\bm\gamma}; {\bm v}_{{\bm\alpha}{\bm\delta}},{\bf V})\varphi({\bm u}; {\bm\alpha}+{\bf X}{\bm\beta}+{\bf Z}{\bm\delta},{\bf R})$, where ${\bf R}= {\bf Z}{\bf D}{\bf Z}^\top+{\bf W}^{-1}$, ${\bf W}={\bf W}_{\bm\eta}={\rm diag}(w_1,\dots,w_n)$ with $w_i=w_{i{\bm\eta}}=b''(\eta_i)$, ${\bm v}_{{\bm\alpha}{\bm\delta}}={\bm\delta}-{\bf D}{\bf Z}^\top{\bf R}^{-1}({\bm\alpha}+{\bf Z}{\bm\delta})+{\bf D}{\bf Z}^\top{\bf R}^{-1}({\bm u}-{\bf X}{\bm\beta})$ with ${\bm v}={\bm v}_{{\bm 0}{\bm 0}}= {\bf D}{\bf Z}^\top{\bf R}^{-1}({\bm u}-{\bf X}{\bm\beta})$, and ${\bf V}={\bf D}-{\bf D}{\bf Z}^\top{\bf R}^{-1}{\bf Z}{\bf D}$. 
\end{lem}

Let
\begin{equation}
\label{eq:definition of H(alpha,delta) normal}
\eqalign{
H({\bm\alpha},{\bm\delta})=&{\rm E}\{\log[\varphi({\bm u};{\bm\alpha}+{\bf X}{\bm\beta}+{\bf Z}{\bm\gamma},{\bf W}^{-1})\varphi({\bm\gamma};{\bm\delta},{\bf D})]\}\cr
=&\int_{\mathbb{R}^r}\int_{\mathbb{R}^n} \log[\varphi({\bm u};{\bm\alpha}+{\bf X}{\bm\beta}+{\bf Z}{\bm\gamma},{\bf W}^{-1})\varphi({\bm\gamma};{\bm\delta},{\bf D})]f({\bm y}|{\bm\gamma})\pi({\bm\gamma})\nu(d{\bm y})d{\bm\gamma}\cr
=&\int_{\mathbb{R}^n}\int_{\mathbb{R}^r} \log[\varphi({\bm\gamma}; {\bm v}_{{\bm\alpha}{\bm\delta}},{\bf V})\varphi({\bm u}; {\bm\alpha}+{\bf X}{\bm\beta}+{\bf Z}{\bm\delta},{\bf R})]  q({\bm\gamma}|{\bm y})\bar f({\bm y})d{\bm\gamma}\nu(d{\bm y})\cr
}
\end{equation}
with $H_1({\bm\alpha})=H({\bm\alpha},{\bm 0})$ and $H_2({\bm\delta})=H({\bm 0},{\bm\delta})$. By ${\bm\eta}={\bf X}{\bm\beta}+{\bf Z}{\bm\gamma}$, we obtain the $i$th component of the gradient vector $\dot H_1({\bm\alpha})$ of $H_1({\bm\alpha})$ as
\begin{equation}
\label{eq:first order partial derivation H(alpha) normal}
{\partial H_1({\bm\alpha})\over\partial\alpha_i}={\rm E}[w_i(u_i-\eta_i-\alpha_i)],~ i=1,\dots,n.
\end{equation}
If $\alpha_i=0$, then the right-hand side of~\eqref{eq:first order partial derivation H(alpha) normal} becomes ${\rm E}[w_i(u_i-\eta_i)]={\rm E}[y_i-b'(\eta_i)]$, leading to $\dot H_1({\bm 0})=\dot h_1({\bm 0})$. For the Hessian matrix $\ddot H_1({\bm\alpha})$ of $H_1({\bm\alpha})$, we have 
\begin{equation}
\label{eq:second order partial derivation H(alpha) normal}
{\partial^2 H_1({\bm\alpha})\over\partial\alpha_i^2}=-{\rm E}(w_i)=-{\rm E}[b''(\eta_i)],~i=1,\dots,n
\end{equation}
with $\partial^2 H_1({\bm\alpha})/\partial\alpha_i\partial\alpha_j=0$ for any distinct $i,j\in\{1,\dots,n\}$, leading to $\ddot H_1({\bm 0})=\ddot h({\bm 0})$. Note that $\ddot H_1({\bm\alpha})$ is negative definite, ${\bm\alpha}={\bm 0}$ is the unique maximizer of $H_1({\bm\alpha})$. Optimizations of $h({\bm\alpha})$ and $H_1({\bm\alpha})$ are equivalent. We extend the conclusion for $H({\bm\alpha},{\bm\delta})$, where ${\bm\alpha}$ and ${\bm\delta}$ can be arbitrary functions of ${\bm y}$, ${\bf X}$, ${\bf Z}$, and ${\bm\gamma}$. 

\begin{lem}
\label{lem:optimization of H(alpha,delta) general}
$H({\bm\alpha},{\bm\delta})\le H({\bm0},{\bm 0})$ and the equality can only be reached by ${\bm\alpha}\stackrel{a.s}={\bm 0}$ and ${\bm\delta}\stackrel{a.s}={\bm 0}$.
\end{lem}

Let ${\bm\alpha}=-{\bf Z}{\bm\delta}$ in~\eqref{eq:definition of H(alpha,delta) normal}. We obtain
\begin{equation}
\label{eq:spacial case H(alpha,delta) normal}
\eqalign{
H(-{\bf Z}{\bm\delta},{\bm\delta})
=&\int_{\mathbb{R}^n}\int_{\mathbb{R}^r} \log[\varphi({\bm\gamma}; {\bm v}_{\bm\delta},{\bf V})\varphi({\bm u};{\bf X}{\bm\beta},{\bf R})]  q({\bm\gamma}|{\bm y})\bar f({\bm y})d{\bm\gamma}\nu(d{\bm y}),\cr
}
\end{equation}
where ${\bm v}_{\bm\delta}={\bm\delta}+{\bf D}{\bf Z}^\top{\bf R}^{-1}({\bm u}-{\bf X}{\bm\beta})={\bm\delta}+{\bm v}$. Because $\varphi({\bm u};{\bf X}{\bm\beta},{\bf R})$ does not vary with ${\bm\delta}$, optimization of $H(-{\bf Z}{\bm\delta},{\bm\delta})$ is equivalent to optimization of
\begin{equation}
\label{eq:spacial case psi(alpha,delta) normal}
\eqalign{
\psi({\bm\delta})=&\int_{\mathbb{R}^n}\int_{\mathbb{R}^r} \log[\varphi({\bm\gamma}; {\bm v}+{\bm\delta},{\bf V})]  q({\bm\gamma}|{\bm y})\bar f({\bm y})d{\bm\gamma}\nu(d{\bm y}).\cr
}
\end{equation}
Be Lemma~\ref{lem:optimization of H(alpha,delta) general}, $\psi({\bm\delta})$ is uniquely maximized at ${\bm\delta}={\bm 0}$ almost surely. By
\begin{equation}
\label{eq:the normal likelihood loss for poster mean and covariance GLMM}
\eqalign{
\log[\varphi({\bm\gamma}; {\bm v}+{\bm\delta},{\bf V})]=-{r\over 2}\log(2\pi)-{1\over 2}\log{\bf V}-{1\over 2}({\bm\gamma}-{\bm\delta}-{\bm v}){\bf V}^{-1}({\bm\gamma}-{\bm\delta}-{\bm v}).
}
\end{equation}
The formulations of ${\bm v}$ and ${\bf V}$ can be used to compute ${\bm\xi}$ and ${\bm\Xi}$, respectively. The only issue is that ${\bm v}$ and ${\bf V}$ depend on ${\bm\gamma}$, implying that they are not identical to ${\bm\xi}$ and ${\bm\Xi}$, respectively. However, we can use the formulations of ${\bm v}$ and ${\bf V}$ to devise an optimization problem to numerically compute the exact values of ${\bm\xi}$ and ${\bm\Xi}$, leading to the main theorem below.

\begin{thm}
\label{thm:main theorem posterior mean and covariance}
(Main Theorem). ${\bm\xi}={\bm v}_{\bm\xi}={\bf D}{\bf Z}^\top{\bf R}_{\bm\xi}^{-1}({\bm u}_{\bm\xi}-{\bf X}{\bm\beta})$ and ${\bm\Xi}={\bf V}_{\bm\xi}={\bf D}-{\bf D}{\bf Z}^\top{\bf R}_{\bm\xi}^{-1}{\bf Z}{\bf D}$, where ${\bm u}_{\bm\xi}={\bm\eta}_{\bm\xi}+[{\bm y}-b'({\bm\eta}_{\bm\xi})]/b''({\bm\eta}_{\bm\xi})$, ${\bf R}_{\bm\xi}={\bf Z}{\bf D}{\bf Z}^\top+{\bf W}_{\bm\xi}^{-1}$, ${\bf W}_{\bm\xi}={\rm diag}[b''({\bm\eta}_{\bm\xi})]$,  and ${\bm\eta}_{\bm\xi}={\bf X}{\bm\beta}+{\bf Z}{\bm\xi}$.
\end{thm}

We devise an algorithm for ${\bm\xi}$ and ${\bm\Xi}$ based on Theorem~\ref{thm:main theorem posterior mean and covariance}. If ${\bm\xi}^{(t)}$ is the $t$th iterated value of ${\bm\xi}$, then ${\bm\Xi}^{(t)}={\bf D}-{\bf D}{\bf Z}^\top{\bf R}_{{\bm\xi}^{(t)}}{\bf Z}{\bf D}$ is the $t$th iterated value of ${\bm\Xi}$. Similarly, ${\bm\eta}_{{\bm\xi}^{(t)}}$, ${\bm\mu}_{{\bm\xi}^{(t)}}$, ${\bm u}_{{\bm\xi}^{(t)}}$, ${\bf W}_{{\bm\xi}^{(t)}}$, and ${\bf R}_{{\bm\xi}^{(t)}}$ are the $t$th iterated values of ${\bm\eta}_{\bm\xi}$, ${\bm\mu}_{\bm\xi}$, ${\bm u}_{\bm\xi}$, ${\bf W}_{\bm\xi}$, and ${\bf R}_{{\bm\xi}}$, respectively. To start the algorithm, we need an initial ${\bm\eta}^{(0)}$. We use it to compute the initial ${\bm\mu}^{(0)}$, ${\bm u}^{(0)}$, ${\bf W}^{(0)}$, and ${\bf R}^{(0)}$, leading to the initial ${\bm\xi}^{(0)}$ and ${\bm\Xi}_{{\bm\xi}^{(0)}}$. We find that ${\bm\eta}^{(0)}$ can be selected by the initial linear component adopted in the iterative reweighted least squares (IRWLS) for the MLE of the GLM~\cite[e.g., Section 4.6]{agresti2002}. We propose Algorithm~\ref{alg:numerical algorithm for posterior mean and covariance}.


\begin{algorithm}
\caption{\label{alg:numerical algorithm for posterior mean and covariance} Special Integral Computation (SIC) for ${\bm\xi}$ and ${\bm\Xi}$}
\begin{algorithmic}[1]
\Statex{{\bf Input}: Data from the GLMM jointly defined by~\eqref{eq:exponential family distribution},~\eqref{eq:generalized linear mixed model}, and~\eqref{eq:distribution of random effects} with ${\bm\beta}$ and ${\bm\omega}$ derived by an estimation method}
\Statex{{\bf Output}: Posterior mean ${\bm\xi}$ and covariance ${\bm\Xi}$}
\Statex{\it Initialization}
\State{Choose an initial ${\bm\eta}^{(0)}$}
\State{Set ${\bm\mu}^{(0)}=b'({\bm\eta}^{(0)})$, ${\bm u}^{(0)}={\bm\eta}^{(0)}+[{\bm y}-b'({\bm\eta}^{(0)})]/b''({\bm\eta}^{(0)})$, ${\bf W}^{(0)}={\rm diag}[b''({\bm\eta}^{(0)})]$, and ${\bf R}^{(0)}=\{{\bf W}^{(0)}\}^{-1}+{\bf Z}{\bf D}{\bf Z}^\top$}
\State{${\bm\xi}^{(0)}\leftarrow{\bf D}{\bf Z}^\top\{{\bf R}^{(0)}\}^{-1}({\bm u}^{(0)}-{\bf X}{\bm\beta})$}
\State{${\bm\eta}_{{\bm\xi}^{(0)}}\leftarrow{\bf X}{\bm\beta}+{\bf Z}{\bm\xi}^{(0)}$,  ${\bm\mu}_{{\bm\xi}^{(0)}}\leftarrow b'({\bm\eta}_{{\bm\xi}^{(0)}})$, ${\bm u}_{{\bm\xi}^{(0)}}\leftarrow{\bm\eta}_{{\bm\xi}^{(0)}}+[{\bm y}-b'({\bm\eta}_{{\bm\xi}^{(0)}})]/b''({\bm\eta}_{{\bm\xi}^{(0)}})$, ${\bf W}_{{\bm\xi}^{(0)}}\leftarrow {\rm diag}[b''({\bm\eta}_{{\bm\xi}^{(0)}})]$,  ${\bf R}_{{\bm\xi}^{(0)}}=\{{\bf W}_{{\bm\xi}^{(0)}}\}^{-1}+{\bf Z}{\bf D}{\bf Z}^\top$, and  ${\bm\Xi}_{{\bm\xi}^{(0)}}\leftarrow{\bf D}-{\bf D}{\bf Z}^\top\{{\bf R}_{{\bm\xi}^{(0)}}\}^{-1}{\bf Z}{\bf D}$ }
\Statex{\it Begin Iteration}
\State{${\bm\xi}^{(t)}\leftarrow{\bf D}{\bf Z}^\top\{{\bf R}_{{\bm\xi}^{(t-1)}}\}^{-1}({\bm u}_{{\bm\xi}^{(t-1)}}-{\bf X}{\bm\beta})$}
\State{${\bm\eta}_{{\bm\xi}^{(t)}}\leftarrow{\bf X}{\bm\beta}+{\bf Z}{\bm\xi}^{(t)}$,  ${\bm\mu}_{{\bm\xi}^{(t)}}\leftarrow b'({\bm\eta}_{{\bm\xi}^{(t)}})$, ${\bm u}_{{\bm\xi}^{(t)}}\leftarrow{\bm\eta}_{{\bm\xi}^{(t)}}+[{\bm y}-b'({\bm\eta}_{{\bm\xi}^{(t)}})]/b''({\bm\eta}_{{\bm\xi}^{(t)}})$, ${\bf W}_{{\bm\xi}^{(t)}}\leftarrow {\rm diag}[b''({\bm\eta}_{{\bm\xi}^{(t)}})]$, ${\bf R}_{{\bm\xi}^{(t)}}=\{{\bf W}_{{\bm\xi}^{(t)}}\}^{-1}+{\bf Z}{\bf D}{\bf Z}^\top$, and  ${\bm\Xi}_{{\bm\xi}^{(t)}}\leftarrow{\bf D}-{\bf D}{\bf Z}^\top\{{\bf R}_{{\bm\xi}^{(t)}}\}^{-1}{\bf Z}{\bf D}$ }
\Statex{\it End Iteration}
\State {Output the final ${\bm\xi}^{(t)}$ and ${\bm\Xi}_{{\bm\xi}^{(t)}}$ as anwers of ${\bm\xi}$ and ${\bm\Xi}$, respectively}
\end{algorithmic}
\end{algorithm}

\begin{cor}
\label{cor:posterior mean and covariance computation exact answers}
The final answers of Algorithm~\ref{alg:numerical algorithm for posterior mean and covariance} are the exact values of ${\bm\xi}$ and ${\bm\Xi}$, respectively.
\end{cor}

The proposed SIC given by Algorithm~\ref{alg:numerical algorithm for posterior mean and covariance} computes the exact ${\bm\xi}$ and ${\bm\Xi}$ based on estimators of ${\bm\beta}$ and ${\bm\omega}$. It can be combined with arbitrary estimation methods for ${\bm\beta}$ and ${\bm\omega}$. The algorithm is used after the estimates of ${\bm\beta}$ and ${\bm\omega}$ are derived. No matter which estimation method is used, the mathematical formulations for the derivation of ${\bm\xi}$ and ${\bm\Xi}$ do not change. The estimation methods for ${\bm\beta}$ and ${\bm\omega}$ significantly affect the properties of the final answers of ${\bm\xi}$ and ${\bm\Xi}$. This is evaluated in our numerical studies.  

\section{Prediction}
\label{sec:prediction}

The SIC also predicts the random effects. In practice, prediction of the random effects is important when GLMMs are used for spatial data. It can be achieved by modifying Step 5 of Algorithm~\ref{alg:numerical algorithm for posterior mean and covariance}. Suppose that spatial data with $n$ observed sites  and $n^*$ unobserved sites are studied. The coordinates of the observed sties are ${\bm s}_1,\dots,{\bm s}_n$. The coordinates of the unobserved sites are ${\bm s}_{1}^*,\dots,{\bm s}_{n^*}^*$. The response and explanatory variables are available at the observed sites. Only the explanatory variables are available at the unobserved sites. A goal of spatial statistics is to predict the random effects and the response at the unobserved sites.

We use ${\bm y}$ and ${\bf X}$ to denote the response vector and the design matrix for the fixed effects at the observed sites, respectively. We use  ${\bm y}^*=(y_1^*,\dots,y_{n^*}^*)^\top$ and ${\bf X}^*=({\bm x}_1^{*\top},\dots,{\bm x}_{n^*}^{*\top})^\top$ to denote those at the unobserved site, respectively. A spatial GLMM (SGLMM) is derived by a modification of~\eqref{eq:exponential family distribution}, \eqref{eq:generalized linear mixed model}, and \eqref{eq:distribution of random effects} with the PDF/PMF as 
\begin{equation}
\label{eq:conditional PMF poisson spatial}
f({\bm y},{\bm y}^*|{\bm\gamma},{\bm\gamma}^*)=\exp\{[{\bm y}^\top{\bm\eta}-{\bm 1}^\top b({\bm\eta})+{\bm 1}^\top c({\bm y})]+[{\bm y}^{*\top}{\bm\eta}^*-{\bm 1}^\top b({\bm\eta}^*)+{\bm 1}^\top c({\bm y}^*)]\},
\end{equation}
 the linear components as
\begin{equation}
\label{eq:linear component of a spatial GLMM}
\left(\begin{array}{c} {\bm\eta}\cr {\bm\eta}^*\cr \end{array}\right)=\left(\begin{array}{c}{\bf X}\cr {\bf X}^*\cr \end{array}\right){\bm\beta}+\left(\begin{array}{c} {\bm\gamma}\cr {\bm\gamma}^*\cr \end{array}\right),
\end{equation}
and the prior for the spatial random effects as 
\begin{equation}
\label{eq:prior for random effects of a spatial GLMM}
\left(\begin{array}{c} {\bm\gamma}\cr {\bm\gamma}^*\cr \end{array}\right)\sim{\mathcal N}\left[\left(\begin{array}{c} {\bm 0}\cr {\bm 0}\cr \end{array}\right), \left(\begin{array}{cc}  {\bf D}_{11} &{\bf D}_{12} \cr {\bf D}_{21} & {\bf D}_{22} \end{array}\right)\right]
\end{equation}
satisfying ${\bf D}_{21}={\bf D}_{12}^\top$. A parametric model specifies the covariance matrix of the spatial random effects between the sites. An example is the Mat\'ern family as
\begin{equation}
\label{eq:matern family}
M_{\bm\omega}( d )={\omega_1\over 1-\omega_1}{(\omega_2d)^{\omega_3}\over 2^{\omega_3-1}\Gamma(\omega_3)}K_{\omega_3}(\omega_2d),
\end{equation}
where ${\bm\omega}=(\omega_1,\omega_2,\omega_3)^\top$ with $\omega_1\in(0,1)$, $\omega_2>0$, and $\omega_3>0$ is the hyperparameter vector, $d$ is the distance between the sites, $K_{\omega_3}(\cdot)$ is the modified Bessel function of the second kind. In the Mat\'ern family, $\omega_1/(1-\omega_1)$, $1/\omega_2$, and $\omega_3$ are the variance, scale, and smoothness parameters, respectively. The Mat\'ern family is isotropic in space. It contains the exponential covariance function as a special case when $\omega_3=0.5$. The Mat\'ern family was first proposed by~\cite{matern1986} and has received more attention since the theoretical work of~\cite{handcock1993,stein1999}. 

We modify Algorithm~\ref{alg:numerical algorithm for posterior mean and covariance} for predicting ${\bm\gamma}^*$ and ${\bm u}^*$ at the unobserved sites. Let ${\bm\xi}^{(t)}$ and ${\bm\xi}^{*(t)}$ be the prediction of ${\bm\gamma}$ and ${\bm\gamma}^*$ in the $t$th iteration of the algorithm. Then, ${\bm\eta}_{{\bm\xi}^{(t)}}={\bf X}{\bm\beta}+{\bm\xi}^{(t)}$ and ${\bm\eta}_{{\bm\xi}^{*(t)}}={\bf X}{\bm\beta}+{\bm\xi}^{*(t)}$ are the prediction of the linear components at the observed and unobserved sites, respectively. It leads to ${\bm\mu}_{{\bm\xi}^{(t)}}=b'({\bm\eta}_{{\bm\xi}^{(t)}})$, ${\bm u}_{{\bm\xi}^{(t)}}={\bm\eta}_{{\bm\xi}^{(t)}}+[{\bm y}-b'({\bm\eta}_{{\bm\xi}^{(t)}})]/b''({\bm\eta}_{{\bm\xi}^{(t)}})$, and ${\bf W}_{{\bm\xi}^{(t)}}= {\rm diag}[b''({\bm\eta}_{{\bm\xi}^{(t)}})]$ for the observed sites,  and ${\bm\mu}_{{\bm\xi}^{*(t)}}=b'({\bm\eta}_{{\bm\xi}^{*(t)}})$, ${\bm u}_{{\bm\xi}^{*(t)}}={\bm\eta}_{{\bm\xi}^{*(t)}}+[{\bm y}-b'({\bm\eta}_{{\bm\xi}^{*(t)}})]/b''({\bm\eta}_{{\bm\xi}^{*(t)}})$, and ${\bf W}_{{\bm\xi}^{*(t)}}= {\rm diag}[b''({\bm\eta}_{{\bm\xi}^{*(t)}})]$ for the unobserved sites. By this idea, we have
\begin{equation}
\label{eq:covariance matrix prediction}
{\bf R}^{(t)}=\left( \begin{array}{cc}  {\bf R}_{11}^{(t)} & {\bf R}_{12}^{(t)} \cr {\bf R}_{21}^{\top} &{\bf R}_{22}^{(t)}\cr \end{array} \right),
\end{equation}
where ${\bf R}_{11}^{(t)}= \{{\bf W}_{{\bm\xi}^{(t)}}\}^{-1}+{\bf D}_{11}$, ${\bf R}_{12}^{(t)}={\bf R}_{21}^{(t)\top}={\bf D}_{12}$, and ${\bf R}_{22}^{(t)}= \{{\bf W}_{{\bm\xi}^{*(t)}}\}^{-1}+{\bf D}_{22}$. There are ${\bm\xi}^{(t+1)}={\bf D}_{11}\{{\bf R}_{11}^{(t)}\}^{-1}({\bm u}_{{\bm\xi}^{(t)}}-{\bf X}{\bm\beta})$ and ${\bm\xi}^{*(t+1)}={\bf R}_{21}^{(t)}\{{\bf R}_{11}^{(t)}\}^{-1}({\bm u}_{{\bm\xi}^{*(t+1)}}-{\bf X}^*{\bm\beta})$ in the $(t+1)$th iteration. The approach to Step 5 of Algorithm~\ref{alg:numerical algorithm for posterior mean and covariance} can be used, leading to the iterations for the prediction. We predict the random effects at the unobserved sites by $\hat{\bm\gamma}^*={\bm\xi}^*={\rm E}({\bm\gamma}^*|{\bm y})$ with ${\bm\xi}^*$ as the final answer of ${\bm\xi}^{*(t)}$. We predict ${\bm y}^*$ by $\hat{\bm y}^*=b'({\bm\eta}_{{\bm\xi}^*})$ and ${\bm u}^*$ by $\hat{\bm u}^*={\bm\eta}_{{\bm\xi}^*}+[{\bm y}-b'({\bm\eta}_{{\bm\xi}^*})]/b''({\bm\eta}_{{\bm\xi}^*})$. They are used in our numerical studies. 

\section{Example}
\label{sec:example}

We specify our SIC to two well-known GLMMs. The first is the binomial GLMM in Section~\ref{subsec:binomial GLMM}.  The second is Poisson GLMM in Section~\ref{subsec:poisson GLMM}. We compute the exact posterior mean ${\bm\xi}$ and covairance ${\bm\Xi}$ in both. The computation does not need the posterior distribution. The computation of the posterior distribution remains unsolved.

\subsection{Binomial GLMM}
\label{subsec:binomial GLMM}

A binomial GLMM assumes $y_i|{\bm\gamma}\sim Bin(m_i,\pi_i)$ independently all $i\in\{1,\dots,n\}$. If the logistic link is used, then the model is
\begin{equation}
\label{eq:logistic in binomial}
\log{{\bm \pi}\over {\bm 1}-{\bm\pi}}={\bm\eta}={\bf X}^\top{\bm\beta}+{\bf Z}^\top{\bm\gamma},
\end{equation}
where ${\bm\pi}=(\pi_1,\dots,\pi_n)^\top$. We choose ${\bm\eta}^{(0)}$ by $\eta_i^{(0)}=\log[(y_i+0.5)/(m_i-y_i+0.5)]$, ${\bf W}^{(0)}$ by $w_i^{(0)}=m_i(y_i+0.5)(m_i-y_i+0.5)/(m_i+1)^2$, and $ u_i^{(0)}=\eta_i^{(0)}+(y_i-n\pi_i^{(0)})/[m_i\pi_i^{(0)}(1-\pi_i^{(0)})]$, where $\pi_i^{(0)}=\exp(\eta_i^{(0)})/[1+\exp(\eta_i^{(0)})]$ for $i=1,\dots,n$. By ${\bf R}^{(0)}={\bf W}^{(0)}+{\bf Z}{\bf D}{\bf Z}^\top$ and ${\bm\xi}^{(0)}={\bf D}{\bf Z}^\top\{{\bf R}^{(0)}\}^{-1}({\bm u}^{(0)}-{\bf X}{\bm\beta})$, we have ${\bm\eta}_{{\bm\xi}^{(0)}}={\bf X}{\bm\beta}+{\bf Z}{\bm\xi}^{(0)}$, ${\bm\pi}_{{\bm\xi}^{(0)}}=\exp({\bm\eta}_{{\bm\xi}^{(0)}})/[1+\exp({\bm\eta}_{{\bm\xi}^{(0)}})]$, ${\bm u}_{{\bm\xi}^{(0)}}={\bm\eta}_{{\bm\xi}^{(0)}}+({\bm y}-{\bm m}\circ{\bm\pi}_{{\bm\xi}^{(0)}})/[{\bm m}\circ{\bm\pi}_{{\bm\xi}^{(0)}}\circ({\bm 1}-{\bm\pi}_{{\bm\xi}^{(0)}})]$, ${\bf W}_{{\bm\xi}^{(0)}}= {\bm m}\circ{\bm\mu}_{{\bm\pi}^{(0)}}\circ(1-{\bm\pi}_{{\bm\xi}^{(0)}})$, ${\bf R}_{{\bm\xi}^{(0)}}=\{ {\bf W}_{{\bm\xi}^{(0)}}  \}^{-1}+{\bf Z}{\bf D}{\bf Z}^\top$, and ${\bm\Xi}_{{\bm\xi}^{(0)}}={\bf D}-{\bf D}{\bf Z}^\top\{{\bf R}_{{\bm\xi}^{(0)}}\}^{-1}{\bf Z}{\bf D}$, where ${\bm m}=(m_1,\dots,m_n)^\top$ and $\circ$ represents the Hadamard (i.e., elementwise) product of vectors or matrices. We obtain the initialization stage. In the iteration stage, there is ${\bm\xi}^{(t)}={\bf D}{\bf Z}^\top\{{\bf R}_{{\bm\xi}^{(t-1)}}\}^{-1}({\bm u}_{{\bm\xi}^{(t-1)}}-{\bf X}{\bm\beta})$. We have ${\bm\eta}_{{\bm\xi}^{(t)}}={\bf X}{\bm\beta}+{\bf Z}{\bm\xi}^{(t)}$, ${\bm\pi}_{{\bm\xi}^{(t)}}=\exp({\bm\eta}_{{\bm\xi}^{(t)}})/[{\bm 1}+\exp({\bm\eta}_{{\bm\xi}^{(t)}})]$, ${\bm u}_{{\bm\xi}^{(t)}}={\bm\eta}_{{\bm\xi}^{(t)}}+({\bm y}-{\bm m}\circ{\bm\pi}_{{\bm\xi}^{(t)}})/[{\bm m}\circ{\bm\pi}_{{\bm\xi}^{(t)}}\circ(1-{\bm\pi}_{{\bm\xi}^{(t)}})]$, ${\bf W}_{{\bm\xi}^{(t)}}= {\bm m}\circ{\bm\pi}_{{\bm\xi}^{(t)}}\circ({\bm 1}-{\bm\pi}_{{\bm\xi}^{(t)}})$, ${\bf R}_{{\bm\xi}^{(t)}}=\{ {\bf W}_{{\bm\xi}^{(t)}}  \}^{-1}+{\bf Z}{\bf D}{\bf Z}^\top$, and ${\bm\Xi}_{{\bm\xi}^{(t)}}={\bf D}-{\bf D}{\bf Z}^\top\{{\bf R}_{{\bm\xi}^{(t)}}\}^{-1}{\bf Z}{\bf D}$. We obtain the iteration stage. We complete Algorithm~\ref{alg:numerical algorithm for posterior mean and covariance}. It provides the exact values of ${\bm\xi}$ and ${\bm\Xi}$ in the logistic GLMM given by~\eqref{eq:logistic in binomial}.

\subsection{Poisson GLMM}
\label{subsec:poisson GLMM}

A Poisson GLMM assumes $y_i|{\bm\gamma}\sim\mathcal{P}(\mu_i)$ independently for all $i\in\{1,\cdots,n\}$. If the log link is used, then the model becomes
\begin{equation}
\label{eq:loglinear in poisson}
\log({\bm\mu})={\bm\eta}={\bf X}^\top{\bm\beta}+{\bf Z}^\top{\bm\gamma},
\end{equation}
where ${\bm\mu}=(\mu_1,\dots,\mu_n)^\top$. We choose ${\bm\eta}^{(0)}=\log({\bm y}+0.5)$ and ${\bf W}^{(0)}={\rm diag}({\bm y}+0.5)$.  By ${\bf R}^{(0)}={\bf W}^{(0)}+{\bf Z}{\bf D}{\bf Z}^\top$ and ${\bm\xi}^{(0)}={\bf D}{\bf Z}^\top\{{\bf R}^{(0)}\}^{-1}({\bm u}^{(0)}-{\bf X}{\bm\beta})$, we have ${\bm\eta}_{{\bm\xi}^{(0)}}={\bf X}{\bm\beta}+{\bf Z}{\bm\xi}^{(0)}$, ${\bm\mu}_{{\bm\xi}^{(0)}}=\exp({\bm\eta}_{{\bm\xi}^{(0)}})$, ${\bm u}_{{\bm\xi}^{(0)}}={\bm\eta}_{{\bm\xi}^{(0)}}+({\bm y}-{\bm\mu}_{{\bm\xi}^{(0)}})/{\bm\mu}_{{\bm\xi}^{(0)}}$, ${\bf W}_{{\bm\xi}^{(0)}}={\rm diag}({\bm\mu}_{{\bm\xi}^{(0)}})$, ${\bf R}_{{\bm\xi}^{(0)}}=\{ {\bf W}_{{\bm\xi}^{(0)}}  \}^{-1}+{\bf Z}{\bf D}{\bf Z}^\top$, and ${\bm\Xi}_{{\bm\xi}^{(0)}}={\bf D}-{\bf D}{\bf Z}^\top\{{\bf R}_{{\bm\xi}^{(0)}}\}^{-1}{\bf Z}{\bf D}$. We obtain the initialization stage. In the iteration stage, there is  ${\bm\xi}^{(t)}={\bf D}{\bf Z}^\top\{{\bf R}_{{\bm\xi}^{(t-1)}}\}^{-1}({\bm u}_{{\bm\xi}^{(t-1)}}-{\bf X}{\bm\beta})$. We have ${\bm\eta}_{{\bm\xi}^{(t)}}={\bf X}{\bm\beta}+{\bf Z}{\bm\xi}^{(t)}$, ${\bm\mu}_{{\bm\xi}^{(t)}}=\exp({\bm\eta}_{{\bm\xi}^{(t)}})$, ${\bm u}_{{\bm\xi}^{(t)}}={\bm\eta}_{{\bm\xi}^{(t)}}+({\bm y}-{\bm\mu}_{{\bm\xi}^{(t)}})/{\bm\mu}_{{\bm\xi}^{(t)}}$, ${\bf W}_{{\bm\xi}^{(t)}}={\rm diag}({\bm\mu}_{{\bm\xi}^{(t)}})$, ${\bf R}_{{\bm\xi}^{(t)}}=\{ {\bf W}_{{\bm\xi}^{(t)}}  \}^{-1}+{\bf Z}{\bf D}{\bf Z}^\top$, and ${\bm\Xi}_{{\bm\xi}^{(t)}}={\bf D}-{\bf D}{\bf Z}^\top\{{\bf R}_{{\bm\xi}^{(t)}}\}^{-1}{\bf Z}{\bf D}$. We obtain the iteration stage. We complete Algorithm~\ref{alg:numerical algorithm for posterior mean and covariance}. It provides the exact values of ${\bm\xi}$ and ${\bm\Xi}$ in the loglinear GLMM given by~\eqref{eq:loglinear in poisson}.

\section{Simulation}
\label{sec:simulation}

We evaluate the performance of our proposed method via Monte Carlo simulations with $1000$ replications. We consider three scenarios. The {\it Oracle scenario} assumes that the true parameters and the random effects are known. It is treated as the ground truth in the evaluation of our method. This approach was first introduced by~\cite{fanli2001} for variable selection problems. It supposes that an Oracle knows the ground truth and can also use the ground truth in the computation. The goal is to provide the optimal situation that a statistical method can reach. The {\it true parameter scenario} is a weak version of the Oracle scenario. It assumes that the true parameters are known but the random effects are not. It uses the true parameters in the computation. The goal is to study the ideal case of Algorithm~\ref{alg:numerical algorithm for posterior mean and covariance}. The {\it parameter estimation scenario} assumes that neither the true parameters (including the hyperparameters) nor the random effects are known. Parameter estimation has to be used before applying our method. The goal is to evaluate the influences of various estimation methods on the properties of our method.

We assumed that the GLMM jointly defined by~\eqref{eq:exponential family distribution}, \eqref{eq:generalized linear mixed model}, and~\eqref{eq:distribution of random effects} was the SGLMM specified by~\eqref{eq:conditional PMF poisson spatial} and~\eqref{eq:linear component of a spatial GLMM}, and~\eqref{eq:prior for random effects of a spatial GLMM} in Section~\ref{sec:prediction}. We fixed $p=2$ and $n=n^*=400$, leading to ${\bf X}=({\bm 1},{\bm x})$ and ${\bf X}^*=({\bm 1}, {\bm x}^*)$ with ${\bm x}\in\mathbb{R}^{400}$ and ${\bm x}^*\in\mathbb{R}^{400}$ in the simulation settings. For each selected ${\bm\beta}=(\beta_0,\beta_1)^\top$ with $\beta_0=8.0$ and $\beta_1=0.0,1.0,2.0,3.0$ respectively, we independently generated the components of ${\bm x}$ and ${\bm x}^*$ from ${\mathcal N}(0,1)$. We computed the fixed effect components by $\beta_0{\bm 1}+\beta_1{\bm x}$ and $\beta_0{\bm 1}+\beta_1{\bm x}^*$ for the observed and unobserved sites respectively. We generated the random effects components ${\bm\gamma}$ at the observed sites and ${\bm\gamma}^*$ at the unobserved sites by~\eqref{eq:prior for random effects of a spatial GLMM} under~\eqref{eq:matern family} with $\omega_1=0.5$, $\omega_2=1.0$, and $\omega_3=0.5$, leading to $M_{\bm\omega}(d)=\exp(-d)$ as the exponential covariance function between the sites. We computed the linear components at the observed and unobserved sites by ${\bm\eta}=\beta_0{\bm 1}+{\bm x}\beta_1+{\bm\gamma}$ and ${\bm\eta}^*=\beta_0{\bm 1}+{\bm x}^*\beta_1+{\bm\gamma}^*$ respectively. We generated the responses by ${\bm y}|{\bm\gamma}\sim {\mathcal P}[\exp(\eta)]$ and ${\bm y}^*|{\bm\gamma}^*\sim{\mathcal P}[\exp(\eta^*)]$ conditionally independently given the random effects respectively. 

After data were generated, we implemented the Oracle, true parameter, and parameter estimation scenarios. In all of those, we considered the predictions of the random effects at the observed and unobserved sites, denoted as $\hat{\bm\gamma}$ and $\hat{\bm\gamma}^*$ respectively. We also investigated the estimates of parameters and hyperparameters, denoted as $\hat{\bm\beta}=(\hat\beta_0,\hat\beta_1)^\top$ and $\hat{\bm\omega}=(\hat\omega_1,\hat\omega_2,0.5)^\top$, respectively. We did not estimate $\omega_3$ because we found that the influence of estimation of $\omega_3$ could be ignored even when it was treated as a hyperparameter. We decided to fix $\omega_3=0.5$ in our comparison.

We used the relative $L_2$-loss to measure the accuracy of the prediction of random effects. The relative-$L_2$ loss at the observed and the unobserved sites were the averages of $RL_2=\|{\bm\gamma}-\hat{\bm\gamma}\|^2/\|{\bm\gamma}\|^2$ $RL_2^*=\|{\bm\gamma}^*-\hat{\bm\gamma}^*\|^2/\|{\bm\gamma}^*\|^2$ in the Monte Carlo simulations. We used the root mean squared error (RMSE) to measure the accuracy of the estimates of the parameters and hyperparameters. They were calculated by the squared root of the sample mean of $\|\beta_0-\hat\beta_0\|^2$, $\|\beta_1-\hat\beta_1\|^2$,  $\|\omega_1-\hat\omega_1\|^2$, and $\|\omega_2-\hat\omega_2\|^2$, denoted as ${\rm RMSE}(\hat\beta_0)$, ${\rm RMSE}(\hat\beta_1)$, ${\rm RMSE}(\hat\omega_1)$, and ${\rm RMSE}(\hat\omega_2)$, respectively. 

The Oracle scenario assumed that the true ${\bm\beta}$, ${\bm\omega}$, and ${\bm\gamma}$ were known and could be used. It predicted ${\bm\gamma}^*$ by
\begin{equation}
\label{eq:predictio of unobserved site Oracle}
\hat{\bm\gamma}_{oracle}^*={\rm E}({\bm\gamma}^*|{\bm\gamma})= {\bf D}_{21}{\bf D}_{11}^{-1}{\bm\gamma}
\end{equation}
at the unobserved sites. Because it used $\hat{\bm\gamma}={\bm\gamma}$, $\hat{\bm\beta}={\bm\beta}$, and $\hat{\bm\omega}={\bm\omega}$, the Oracle scenario satisfied $RL_2^*>0$ and $RL_2={\rm RMSE}(\hat\beta_0)={\rm RMSE}(\hat\beta_1)={\rm RMSE}(\hat\omega_1)={\rm RMSE}(\hat\omega_2)=0$. 

The true parameter scenario assumed that the true ${\bm\beta}$ and ${\bm\omega}$ were known and could be used. It predicted ${\bm\gamma}$ by ${\bm\xi}$ and ${\bm\gamma}^*$ by ${\bm\xi}^*$. The computation was carried out based on the true ${\bm\beta}$ and ${\bm\omega}$ in Algorithm~\ref{alg:numerical algorithm for posterior mean and covariance} for ${\bm\xi}$ and a similar algorithm for ${\bm\xi}^*$. The true parameter scenario satisfied $RL_2,RL_2^*>0$ and ${\rm RMSE}(\hat\beta_0)={\rm RMSE}(\hat\beta_1)={\rm RMSE}(\hat\omega_1)={\rm RMSE}(\hat\omega_2)=0$. 

The parameter estimation scenario assumed that none of ${\bm\beta}$, ${\bm\omega}$, or ${\bm\gamma}$ was known. It estimated ${\bm\beta}$ and ${\bm\omega}$ before computing ${\bm\xi}$ and ${\bm\xi}^*$. The parameter estimation scenario had positive $RL_2$ ,$RL_2^*$, ${\rm RMSE}(\hat\beta_0)$, ${\rm RMSE}(\hat\beta_1)$, ${\rm RMSE}(\hat\omega_1)$, and ${\rm RMSE}(\hat\omega_2)$ values. The performance of these quantities depended on estimation methods.

\begin{table}
\caption{\label{tab:prediction of random effects} Simulation with $1000$ replications for the combinations of our proposed SIC with the previous PM, PQL, LA, and INLA estimation methods. The SIC+True method represents our method based on the true ${\bm\beta}$ and ${\bm\omega}$. The Oracle method represents  the ground truth.}
\begin{center}
\begin{tabular}{ccccccc}\hline
      &  \multicolumn{2}{c}{Prediction}  &  \multicolumn{4}{c}{RMSE of Estimators} \\
Method & $RL_2$ & $RL_2^*$  &  $\hat\beta_0$ & $\hat\beta_1$ & $\hat\omega_1$ & $\hat\omega_2$   \\\hline 
            &  \multicolumn{6}{c}{$\beta_0=8.0$, $\beta_1=0.0$}\\
Oracle & $0$  & $0.501$ &  $0$ & $0$ & $0$ & $0$ \\
SIC+True &  $0.001$ & $0.501$ &  $0$ & $0$ & $0$ & $0$ \\
SIC+PM &  $0.011$ &  $0.512$ &  $0.100$ & $0.026$ & $0.023$ & $0.117$ \\
SIC+PQL &  $0.011$ &  $0.512$ & $0.100$ & $0.026$ & $0.024$ & $0.118$ \\
SIC+LA & $0.011$ & $0.512$ &  $0.100$ & $0.026$ & $0.024$ & $0.117$ \\
SIC+INLA & $19.09$ & $19.77$ & $4.204$ & $0.052$ & $0.214$ & $1.000$ \\
            &  \multicolumn{6}{c}{$\beta_0=8.0$, $\beta_1=1.0$}\\
Oracle & $0$  & $0.502$ &   $0$ & $0$ & $0$ & $0$  \\
SIC+True &  $0.001$ & $0.503$ &  $0$ & $0$ & $0$ & $0$ \\
SIC+PM &  $0.014$ &  $0.517$ &  $0.111$ & $0.027$ & $0.027$ & $0.120$ \\
SIC+PQL &  $0.014$ &  $0.517$ & $0.111$ & $0.027$ & $0.027$ & $0.121$ \\
SIC+LA & $0.014$ & $0.517$ &  $0.112$ & $0.028$ & $0.027$ & $0.121$ \\
SIC+INLA & $18.60$ & $18.65$ & $4.093$ & $0.075$ & $0.223$ & $1.001$ \\
            &  \multicolumn{6}{c}{$\beta_0=8.0$, $\beta_1=2.0$}\\
Oracle & $0$   & $0.500$ &   $0$ & $0$ & $0$ & $0$  \\
SIC+True &  $0.003$ & $0.501$ &  $0$ & $0$ & $0$ & $0$ \\
SIC+PM &  $0.019$ &  $0.517$ &  $0.121$ & $0.032$ & $0.025$ & $0.128$ \\
SIC+PQL &  $0.080$ &  $0.589$ & $0.271$ & $0.034$ & $0.378$ & $0.855$ \\
SIC+LA & $0.079$ & $0.587$ &  $0.267$ & $0.035$ & $0.370$ & $0.840$ \\
SIC+INLA & $19.71$ & $20.27$ & $4.247$ & $0.147$ & $0.220$ & $1.000$ \\
            &  \multicolumn{6}{c}{$\beta_0=8.0$, $\beta_1=3.0$}\\
Oracle & $0$  & $0.504$ &   $0$ & $0$ & $0$ & $0$ \\
SIC+True &  $0.011$ & $0.508$ &  $0$ & $0$ & $0$ & $0$ \\
SIC+PM &  $0.026$ &  $0.523$ &  $0.119$ & $0.028$ & $0.024$ & $0.118$ \\
SIC+PQL &  $0.177$ &  $0.658$ & $0.295$ & $0.217$ & $0.412$ & $0.991$ \\
SIC+LA & $0.172$ & $0.652$ &  $0.285$ & $0.226$ & $0.409$ & $0.887$ \\
SIC+INLA & $20.89$ & $21.17$ & $4.350$ & $0.200$ & $0.209$ & $1.000$ \\\hline
\end{tabular}
\end{center}
\end{table}

We investigated seven previous estimation methods. They were the PM~\cite{zhang2023}, the PQL~\cite{breslow1993}, the LA~\cite{evangelou2011}, the INLA~\cite{rue2009}, the geoRglm~\cite{christensen2002}, the PrevMap~\cite{giorgi2017}, and the geoCount~\cite{jing2015}. In the literature, estimation of parameters and hyperparameters in GLMMs can be classified into likelihood-based or Bayesian methods. In the category of likelihood-based estimation, the PQL is a restricted maximum likelihood (REML) method. The LA and the PM are the maximum likelihood (ML) methods. The PrevMap is a Monte Carlo maximum likelihood (MCML) method. In the category of Bayesian estimation, the INLA is an approximate Bayesian comutation (ABC) method. The geoCount and geoRglm are MCMC methods. We implemented these methods via the corresponding packages of \textsf{R}. The implementation of the PrevMap, geoCount, and geoRglm methods failed due to the presence of the slope in the SGLMM. The implementation succeeded after the slope was removed. We adopted the PM, the PQL, the LA, and the INLA and discarded the PrevMap, the geoCount, and the geoRglm methods in the comparison. 

We generated $1000$ datasets from the SGLMM jointly defined by~\eqref{eq:conditional PMF poisson spatial} and~\eqref{eq:linear component of a spatial GLMM}, and~\eqref{eq:prior for random effects of a spatial GLMM} for each selected $\beta_1$ with $\beta_0=8.0$, $n=n^*=400$, and ${\bm\omega}=(0.5,1.0,0.5)^\top$. We computed the $RL_2$, $RL_2^*$, ${\rm RMSE}(\hat\beta_0)$, ${\rm RMSE}(\hat\beta_1)$, ${\rm RMSE}(\hat\omega_1)$, and ${\rm RMSE}(\hat\omega_2)$ values in our simulation (Table~\ref{tab:prediction of random effects}). The Oracle method was the ground truth. It used ${\bm\gamma}$ to predict ${\bm\gamma}^*$ by~\eqref{eq:predictio of unobserved site Oracle}. All the remaining methods were compared with the ground truth. The true parameter scenario (i.e., SIC+True) did not estimate the parameters or the hyperparameters. It completely reflected the properties of Algorithm~\ref{alg:numerical algorithm for posterior mean and covariance} for ${\bm\xi}$ and a similar algorithm for ${\bm\xi}^*$. The corresponding $RL_2$ value was almost $0$. The corresponding $RL_2^*$ value was close to that of the ground truth. We can conclude that our method is precise and the results are accurate. 

 The parameter estimation scenario estimated $\beta_0$, $\beta_1$, $\omega_1$, and $\omega_2$ before the computation of ${\bm\xi}$ and ${\bm\xi}^*$. It involved four estimation methods. The PM method worked the best because it had the least RMSE values of the estimators. The properties was not significantly affected by the growth of the slope (i.e., $\beta_1$). The performance of the PQL and the LA became worse as the slope increased. The INLA method was imprecise and the results were unreliable. The findings of the parameter estimation scenario suggest to use the SIC+PM method in practice. 

In the end, we evaluated the computational time. The computation of ${\bm\xi}$ and ${\bm\xi}^*$ took less than $10$ seconds, indicating the the implementation of the SIC method is fast. When ${\bm\beta}$ and ${\bm\omega}$ were treated as unknown, the PM method took around $20$ seconds in the computation of $\hat{\bm\beta}$ and $\hat{\bm\omega}$. The PQL and LA methods took around $20$ seconds when $\beta_1=0.0$. The computation increased to many minutes when $\beta_1=3.0$. The INLA method took around $25$ seconds in all the cases. Because it was imprecise, we do not recommend the INLA method to estimate the parameters and hyperparameters. It is better to use the SIC+PM method in real-world applications.

\section{Application}
\label{sec:application}

We applied our method to the Weed data which were initially analyzed by~\cite{guillot2009}. The Weed data were collected from $100$ sites specified by $0.5\times 0.75{\rm m}^2$ frames at the Bjertorp farm with $30$ hectares area in sizes located at $58.26^o{\rm N}$ and $13.13^o{\rm E}$ in southwestern Sweden. Because collecting exact weed counts in agricultural fields is expensive and time-consuming, statistical methods are recommended for predicting weed counts to reduce the cost at an appropriate level. To fulfill the research task, previous work used the spatial Poisson lognormal model as $y_i|\gamma_i\sim{\mathcal P}(\mu_i)$ independently with 
\begin{equation}
\label{eq:intercept SGLMM}
\log\mu_i=\beta_0+\gamma_i
\end{equation}
for $i=1,\dots,100$, where the covariance function between the sites were modeled by the Mat\'ern family given by~\eqref{eq:matern family} with $\omega_3=0.5$. It was pointed out by~\cite{zhang2023} that the usage of $\omega_3$ was appropriate. 

A concern was whether a better result could be derived if the longitude (i.e., Lon) and the latitude (i.e., Lat) were added to the linear components of~\eqref{eq:intercept SGLMM}. The concern arose from an initial analysis based on the comparison among three GLMs derived by excluding the random effects (i.e., $\gamma_i$) from the model. The first was the null model as $\log\mu_i=\beta_0$. The second was the main effect model as $\log\mu_i=\beta_0+\beta_1{\rm Lon}+\beta_2{\rm Lat}$. The third was the quadratic and interaction effects as $\log\mu_i=\beta_0+\beta_1{\rm Lon}+\beta_2{\rm Lat}+\beta_3{\rm Lon}^2+\beta_4{\rm Lat}^2+\beta_5{\rm Lon}\times{\rm Lat}$. We fitted the three models individually. We obtained their residual deviance (i.e., deviance goodness-of-fit statistic) values as $G_1^2=4506.9$, $G_2^2=4439.6$, and $G_3^2=4356.2$, respectively. The likelihood ratio statistics for the significance of the quadratic and interaction effects of the longitude and latitude was $G_2^2-G_3^2=83.4$. The corresponding $p$-value was less than $10^{-15}$ based on the approximate $\chi_3^2$-distribution. The likelihood ratio statistics for the linear effects was $G_1^2-G_2^2=67.4$. The corresponding $p$-value was still less than $10^{-15}$ based on the approximate $\chi_2^2$-distribution. The initial analysis showed that it was important to justify the linear and quadratic effects of the longitude and the latitude in the spatial analysis of the Weed data. 

We considered two SGLMMs for the concern. The first was 
\begin{equation}
\label{eq:linear longitude and latitude SGLMM}
\log\mu_i=\beta_0+\beta_1{\rm Lon}+\beta_2{\rm Lat}+\gamma_i.
\end{equation}
The second was 
\begin{equation}
\label{eq:quadratic longitude and latitude SGLMM}
\log\mu_i=\beta_0+\beta_1{\rm Lon}+\beta_2{\rm Lat}+\beta_3{\rm Lon}^2+\beta_4{\rm Lat}^2+\beta_5{\rm Lon}\times{\rm Lat}+\gamma_i.
\end{equation}
We carried out a validation study to compare~\eqref{eq:intercept SGLMM}, \eqref{eq:linear longitude and latitude SGLMM}, and \eqref{eq:quadratic longitude and latitude SGLMM}. We randomly partitioned the Weed data into training and testing datasets denoted as ${\mathcal A}_{train}$ and ${\mathcal A}_{test}$ with $n_{train}=|{\mathcal A}_{train}|$ and $n_{test}=|{\mathcal A}_{test}|$ observations satisfying $n_{train}+n_{test}=100$, respectively. We treated the sites contained in the training and testing datasets as the observed and unobserved sites, respectively. We used the SIC+PM, SIC+PQL, SIC+LA, and SIC+INLA methods to estimate the parameters and hyperparameters (i.e., based on ${\mathcal A}_{train}$) and predict the responses at the unobserved sites (i.e., based on ${\mathcal A}_{test}$) under the three models, respectively. We computed the corresponding predicted deviance goodness-of-fit as $G_{test}^2=2\sum_{i\in{\mathcal A}_{test}}[y_i^*\log(y_i^*/\hat y_i^*)-(y_i^*-\hat y_i^*)]$, where $y_i^*$ and $\hat y_i^*$ were the observed and predicted counts at the unobserved sites, respectively. We repeated the procedure $1000$ times. We studied various options of $n_{train}$ and $n_{test}$. The main findings did not change. We decided to report our results based on $n_{traing}=80$ and $n_{test}=20$. For the SIC+PM, the corresponding $G_{test}^2$ values were $639$, $619$, and $682$, respectively. For the SIC+PQL, the corresponding $G_{test}^2$ values were $630$, $620$, and $676$, respectively. For the SIC+LA, the corresponding $G_{test}^2$ values were $640$, $620$, and $684$, respectively. We did not get the result of the INLA because it failed in~\eqref{eq:quadratic longitude and latitude SGLMM}. The validation study suggested to use~\eqref{eq:linear longitude and latitude SGLMM}.

\begin{table}
\caption{\label{tab:estimates of the main effect model weed data}Estimates of models parameters and hyperparameters of~\eqref{eq:linear longitude and latitude SGLMM} for the whole Weed data by the PM, PQL, and LA methods.}
\begin{center}
\begin{tabular}{cccccc}\hline
 Method & $\hat\beta_0$ & $\hat\beta_1$ & $\hat\beta_2$ & $\hat\omega_1$ & $\hat\omega_2$ \\\hline
 MP & $4.63$ & $-1.35(10^{-4})$ & $-1.87(10^{-3})$ & $0.456$ & $1.54(10^{-2})$   \\
 PQL & $4.50$ & $3.43(10^{-4})$ & $-1.77(10^{-3})$ & $0.543$ & $9.57(10^{-3})$ \\
 LA & $4.48$ & $1.75(10^{-4})$ & $-1.70(10^{-3})$ & $0.467$ & $1.43(10^{-2})$  \\\hline
\end{tabular}
\end{center}
\end{table}

\begin{figure}
\centering
\includegraphics[angle=270,width=3.4in]{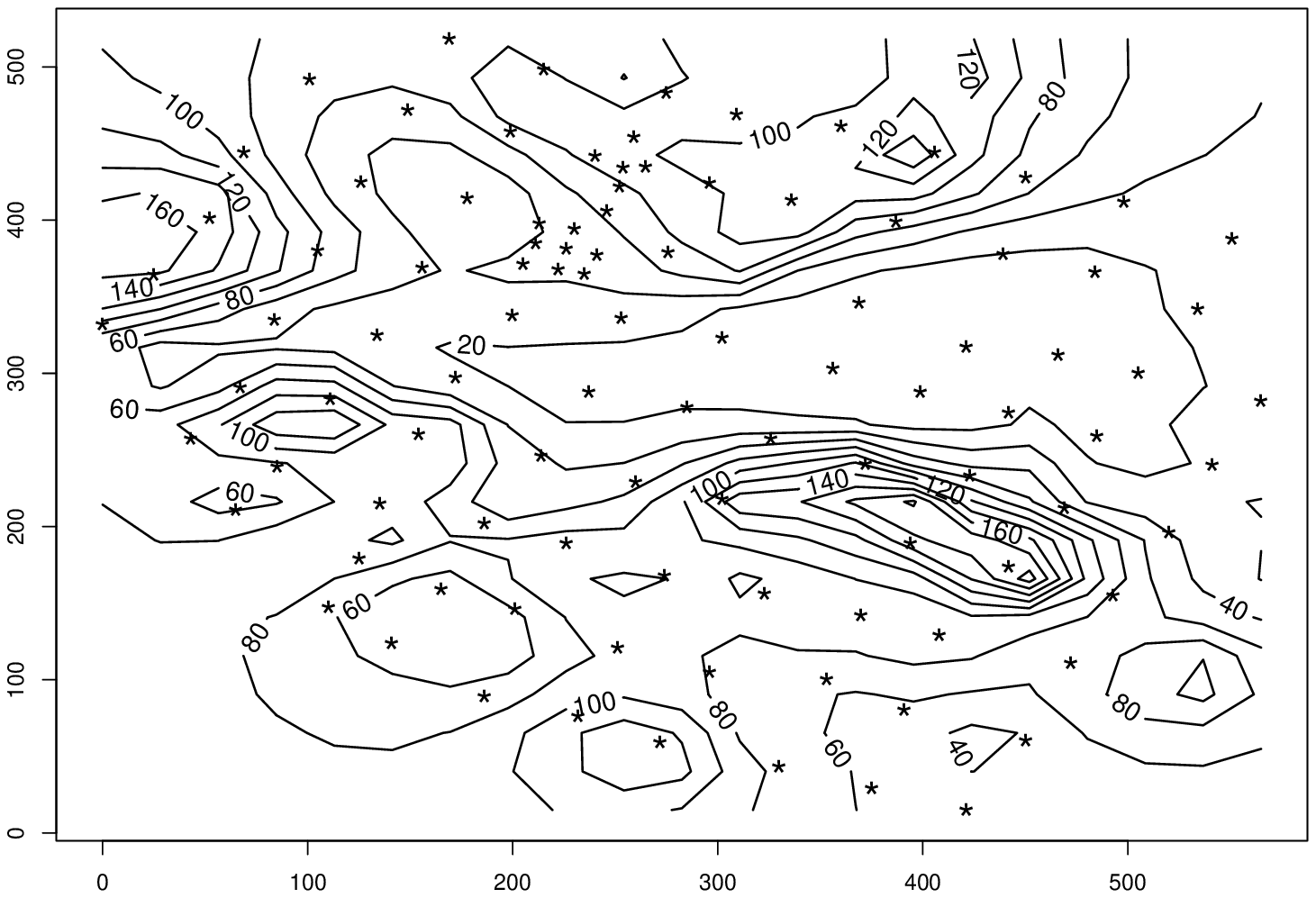}
\caption{\label{fig:prediction of weed counts SIC+PM} Contour plot for predicted weed counts at unobserved sites in the Bjertorp farm by the SIC+PM method, where the locations of samples are marked as *.}
\end{figure}

Based on~\eqref{eq:linear longitude and latitude SGLMM}, we applied the SIC+PM, the SIC+PQL, and the SIC+LA methods to the whole Weed data. The Bjertorp farm had $8\times 10^{5}$ frames within the $30$ hectares area. The Weed data only contained $100$ frames. The goal was the prediction of the weed counts at the remaining frames. The frames contained in the Weed data were treated as the observed sites. The remaining frames were treated as the unobserved sites. We carried out a two-stage procedure to predict the weed counts at the unobserved sites. We estimated the parameters and hyparameters of~\eqref{eq:linear longitude and latitude SGLMM} by the PM, the PQL, and the LA methods in the estimation stage, respectively (Table~\ref{tab:estimates of the main effect model weed data}). The results of the estimation stage were implemented to the prediction stage described in Section \ref{sec:prediction}. It provided the prediction of the weed counts at the unobserved sites. We compared the predicted weed counts from the SIC+PM, the SIC+PQL, and the SIC+LA methods. We found the results were close. We plotted the results obtained from the SIC+PM method, leading to Figure~\ref{fig:prediction of weed counts SIC+PM}. It reflected the pattern of the predicted weed counts at the remaining $799900$ frames not contained in the Weed data. 

The statistical model we found was different from thpse used in the previous work. The previous work focused on the intercept only SGLMM given by~\eqref{eq:intercept SGLMM}~\cite[e.g.]{guillot2009,jing2015,zhang2023}. We found that the main effect SGLMM given by~\eqref{eq:linear longitude and latitude SGLMM} was better. The finding was achieved by a validation study for the Weed data. Because the response follows the Poisson and the prior of the random effects is normal, the posterior distribution of the random effects given the response contains intractable integrals. It is hard to use the posterior distribution in computing the predicted counts at the unobserved sites. The proposed SIC method does not use the posterior distribution. It does not suffer from the computational difficulty caused by intractable integrals, leading to the findings of the application. 

\section{Conclusion}
\label{sec:conclusion}

Intractable integrals are often treated as one of the main long-standing problems in Bayesian statistics under non-conjugate priors. Previous work uses Monte Carlo computations for the posterior distributions. This approach may not work well if the dimension of an integral integral is high. The proposed method does not need a computation of the posterior distribution. It directly computes the posterior mean and covariance without using the posterior distribution. We treat it as an alternative approach for Bayesian statistics under an non-conjugate prior. We investigate a special case of the approach. We point out that high-dimensional integrals are not an issue in the Bayesian inference on GLMMs for count responses when the prior distributions of the random effects are normal. We expect that our idea can be implemented into the general Bayesian framework when intractable integrals are contained in the posterior distributions. This is left to future research.

\appendix

\section{Proofs}
\label{sec:proofs}

\noindent
{\bf Proof of Lemma~\ref{lem:normal density posterior identity general}.}  We express the logarithm of the left-hand side as
$$
\log[\varphi({\bm u};{\bm\alpha}+{\bf X}{\bm\beta}+{\bf Z}{\bm\gamma},{\bf W}^{-1})\varphi({\bm\gamma};{\bm\delta},{\bf D})]=-{n+r\over 2}\log(2\pi)+T_1+T_2+T_3+T_4+T_5+T_6,
$$
where 
$$\eqalign{
T_1=&-(1/2)[\log|\det({\bf W}^{-1})|+\log|\det({\bf D})|],\cr
T_2=&-(1/2)({\bm u}-{\bm\alpha}-{\bf X}{\bm\beta}){\bf W}({\bm u}-{\bm\alpha}-{\bf X}{\bm\beta}),\cr
T_3=& {\bm\gamma}^\top{\bf Z}^\top {\bf W}({\bm u}-{\bm\alpha}-{\bf X}{\bm\beta}),\cr
T_4=&-(1/2){\bm\gamma}^\top({\bf D}^{-1}+{\bf Z}^\top{\bf W}{\bf Z}^\top){\bm\gamma},\cr
T_5=&{\bm\delta}^\top {\bf D}^{-1}{\bm\gamma},\cr
T_6=& -(1/2) {\bm\delta}^\top{\bf D}^{-1}{\bm\delta}.\cr
}$$
 By ${\bm v}_{{\bm\alpha}{\bm\delta}}=({\bf I}-{\bf D}{\bf Z}^\top{\bf R}^{-1}{\bf Z}){\bm\delta}+{\bf D}{\bf Z}^\top{\bf R}^{-1}({\bm u}-{\bm\alpha}-{\bf X}{\bm\beta})$, we express the right-hand side as
$$
\log[\varphi({\bm\gamma}; {\bm v}_{{\bm\alpha}{\bm\delta}},{\bf V})\varphi({\bm u}; {\bm\alpha}+{\bf X}{\bm\beta}+{\bf Z}{\bm\delta},{\bf R})]=-{n+r\over 2}\log(2\pi)+\tilde T_1+\tilde T_2+\tilde T_3+\tilde T_4+\tilde T_5+\tilde T_6,
$$
where 
$$\eqalign{
\tilde T_1=&-(1/2)[\log|\det({\bf R})|+\log|\det({\bf D}-{\bf D}{\bf Z}^\top{\bf R}^{-1}{\bf Z}{\bf D})|],\cr
\tilde T_2=& -(1/2)({\bm u}-{\bm\alpha}-{\bf X}{\bm\beta})^\top[{\bf R}^{-1}-{\bf R}^{-1}{\bf Z}{\bf D}({\bf D}-{\bf D}{\bf Z}^\top{\bf R}^{-1}{\bf Z}{\bf D})^{-1}{\bf D}{\bf Z}^\top{\bf R}^{-1}]({\bm u}-{\bm\alpha}-{\bf X}{\bm\beta}),\cr
\tilde T_3=&{\bm\gamma}^\top({\bf D}-{\bf D}{\bf Z}^\top{\bf R}^{-1}{\bf Z}{\bf D})^{-1}{\bf D}{\bf Z}^\top{\bf R}^{-1} ({\bm u}-{\bm\alpha}-{\bf X}{\bm\beta}),\cr
\tilde T_4= &-(1/2){\bm\gamma}^\top({\bf D}-{\bf D}{\bf Z}^\top{\bf R}^{-1}{\bf Z}{\bf D})^{-1}{\bm\gamma},\cr
\tilde T_5=& {\bm\delta}^\top({\bf I}-{\bf Z}^\top{\bf R}^{-1}{\bf Z}{\bf D})({\bf D}-{\bf D}{\bf Z}^\top{\bf R}^{-1}{\bf Z}{\bf D})^{-1}{\bm\gamma},  \cr
\tilde T_6=&-(1/2) {\bm\delta}^\top[{\bf Z}^\top{\bf R}^{-1}{\bf Z}+({\bf I}-{\bf Z}^\top{\bf R}^{-1}{\bf Z}{\bf D})({\bf D}-{\bf D}{\bf Z}^\top{\bf R}^{-1}{\bf Z}{\bf D})^{-1}({\bf I}-{\bf D}{\bf Z}^\top{\bf R}^{-1}{\bf Z})]{\bm\delta}.\cr
}$$
 We want to show $T_1=\tilde T_1$, $T_2=\tilde T_2$, $T_3=\tilde T_3$, $T_4=\tilde T_4$, $T_5=\tilde T_5$, and $T_6=\tilde T_6$. By the Woodbury matrix identity, we obtain $({\bf D}^{-1}+{\bf Z}^\top{\bf W}{\bf Z})^{-1}={\bf D}-{\bf D}{\bf Z}^\top({\bf W}^{-1}+{\bf Z}{\bf D}{\bf Z}^\top)^{-1}{\bf Z}{\bf D}$, implying ${\bf D}^{-1}+{\bf Z}^\top{\bf W}{\bf Z}=[{\bf D}-{\bf D}{\bf Z}^\top({\bf W}^{-1}+{\bf Z}{\bf D}{\bf Z}^\top)^{-1}{\bf Z}{\bf D}]^{-1}$, leading to $T_4=\tilde T_4$. Still using the Woodbury matrix identity, we obtain
$$\eqalign{
&{\bf R}^{-1}-{\bf R}^{-1}{\bf Z}{\bf D}({\bf D}-{\bf D}{\bf Z}^\top{\bf R}^{-1}{\bf Z}{\bf D})^{-1}{\bf D}{\bf Z}^\top{\bf R}^{-1}\cr
=&({\bf W}^{-1}+{\bf Z}{\bf D}{\bf Z}^\top)^{-1}+({\bf W}^{-1}+{\bf Z}{\bf D}{\bf Z}^\top)^{-1}{\bf Z}{\bf D}\cr
&\hspace{2cm}[{\bf D}-{\bf D}{\bf Z}^\top({\bf W}^{-1}+{\bf Z}{\bf D}{\bf Z}^\top)^{-1}{\bf Z}{\bf D}]^{-1}{\bf D}{\bf Z}^\top({\bf W}^{-1}+{\bf Z}{\bf D}{\bf Z}^\top)^{-1}\cr
=&({\bf W}^{-1}+{\bf Z}{\bf D}{\bf Z}^\top)^{-1}+({\bf W}^{-1}+{\bf Z}{\bf D}{\bf Z}^\top)^{-1}{\bf Z}{\bf D}{\bf Z}^\top{\bf W}\cr
=&{\bf W},
}$$
leading to $T_2=\tilde T_2$. Using the above, we obtain ${\bf R}^{-1}+{\bf R}^{-1}{\bf Z}{\bf D}{\bf Z}^\top{\bf W}={\bf W}$. Then,
$$\eqalign{
&({\bf I}-{\bf Z}^\top{\bf R}^{-1}{\bf Z}{\bf D})[{\bf D}-{\bf D}{\bf Z}^\top({\bf W}^{-1}+{\bf Z}{\bf D}{\bf Z}^\top)^{-1}{\bf Z}{\bf D}]^{-1}\cr
=& ({\bf I}-{\bf Z}^\top{\bf R}^{-1}{\bf Z}{\bf D})({\bf D}^{-1}+{\bf Z}^\top{\bf W}{\bf Z})\cr
=&{\bf D}^{-1}+{\bf Z}^\top{\bf W}{\bf Z}-{\bf Z}^\top{\bf R}^{-1}{\bf Z}-{\bf Z}^\top{\bf R}^{-1}{\bf Z}{\bf D}{\bf Z}^\top{\bf W}{\bf Z}\cr
=&{\bf D}^{-1}+{\bf Z}^\top{\bf W}{\bf Z}-{\bf Z}^\top( {\bf R}^{-1}-{\bf R}^{-1}{\bf Z}{\bf D}{\bf Z}^\top{\bf W} ){\bf Z}\cr
=&{\bf D}^{-1},
}$$
leading to $T_5=\tilde T_5$. Further, we have
$$\eqalign{
&{\bf Z}^\top{\bf R}^{-1}{\bf Z}+({\bf I}-{\bf Z}^\top{\bf R}^{-1}{\bf Z}{\bf D})({\bf D}-{\bf D}{\bf Z}^\top{\bf R}^{-1}{\bf Z}{\bf D})^{-1}({\bf I}-{\bf D}{\bf Z}^\top{\bf R}^{-1}{\bf Z})\cr
=& {\bf Z}^\top{\bf R}^{-1}{\bf Z}+{\bf D}^{-1}({\bf I}-{\bf D}{\bf Z}^\top{\bf R}^{-1}{\bf Z})\cr
=& {\bf D}^{-1},
}$$
leading to $\tilde T_6=T_6$. By
$$\eqalign{
& ({\bf D}-{\bf D}{\bf Z}^\top{\bf R}^{-1}{\bf Z}{\bf D})^{-1}{\bf D}{\bf Z}^\top{\bf R}^{-1}   \cr
=&[{\bf D}-{\bf D}{\bf Z}^\top({\bf W}^{-1}+{\bf Z}{\bf D}{\bf Z}^\top)^{-1}{\bf Z}{\bf D}]^{-1}{\bf D}{\bf Z}^\top({\bf W}^{-1}+{\bf Z}{\bf D}{\bf Z}^\top)^{-1}\cr
=&({\bf D}^{-1}+{\bf Z}^\top{\bf W}{\bf Z}){\bf D}{\bf Z}^\top({\bf W}^{-1}+{\bf Z}{\bf D}{\bf Z}^\top)^{-1}\cr
=& {\bf Z}^\top ({\bf W}^{-1}+{\bf Z}{\bf D}{\bf Z}^\top)^{-1}+{\bf Z}^\top{\bf W}{\bf Z}{\bf D}{\bf Z}^\top({\bf W}^{-1}+{\bf Z}{\bf D}{\bf Z}^\top)^{-1}\cr
=&{\bf Z}^\top {\bf W}{\bf W}^{-1}({\bf W}^{-1}+{\bf Z}{\bf D}{\bf Z}^\top)^{-1}+{\bf Z}^\top{\bf W}{\bf Z}{\bf D}{\bf Z}^\top({\bf W}^{-1}+{\bf Z}{\bf D}{\bf Z}^\top)^{-1}\cr
=&{\bf Z}^\top{\bf W},
}$$
we obtain $T_3=\tilde T_3$. For $T_1$ and $\tilde T_1$, we examine the determinant of 
$${\bf C}=\left( \begin{array}{cc}{\bf D} & {\bf D} {\bf Z}^\top \cr {\bf Z}{\bf D}  & {\bf W}^{-1}+{\bf Z}{\bf D}{\bf Z}^\top   \end{array}  \right).$$
We have
$$\eqalign{
\det({\bf C})=&\det \left[ \left(  \begin{array}{cc}{\bf D} & {\bf D} {\bf Z}^\top \cr {\bf Z}{\bf D}  & {\bf W}^{-1}+{\bf Z}{\bf D}{\bf Z}^\top   \end{array}  \right)\left(\begin{array}{cc} {\bf I} & -{\bf Z}^\top \cr {\bf 0} & {\bf I} \end{array}   \right) \right]\cr
=&\det\left(\begin{array}{cc}{\bf D} & {\bf 0} \cr {\bf Z}{\bf D} & {\bf D} \end{array}   \right) \cr
=&\det({\bf D})\det({\bf W}^{-1})
}$$
and 
$$\eqalign{
\det({\bf C})=&\det \left[\left(\begin{array}{cc} {\bf I} &  -{\bf D}{\bf Z}^\top({\bf W}^{-1}+{\bf Z}{\bf D}{\bf Z}^\top)^{-1} \cr {\bf 0} & {\bf I} \end{array}   \right)  \left(  \begin{array}{cc}{\bf D} & {\bf D}{\bf Z}^\top \cr {\bf Z}{\bf D}  & {\bf W}^{-1}+{\bf Z}{\bf D}{\bf Z}^\top   \end{array}  \right)\right]\cr
=&\det\left( \begin{array}{cc} {\bf D}-{\bf D}{\bf Z}^\top({\bf W}^{-1}+{\bf Z}{\bf D}{\bf Z}^\top)^{-1}{\bf Z}{\bf D}& {\bf 0} \cr {\bf Z}{\bf D} & {\bf W}^{-1}+{\bf Z}{\bf D}{\bf Z}^\top \cr  \end{array}   \right)\cr
=&\det[ {\bf D}-{\bf D}{\bf Z}^\top({\bf W}^{-1}+{\bf Z}{\bf D}{\bf Z}^\top)^{-1}{\bf Z}{\bf D}]\det({\bf W}^{-1}+{\bf Z}{\bf D}{\bf Z}^\top),
}$$
leading to $T_1=\tilde T_1$.  \qed

\noindent
{\bf Proof of Lemma~\ref{lem:optimization of H(alpha,delta) general}.} The gradient vector of $H_2({\bm\delta})$ with respect to ${\bm\delta}$ is $\dot H_2({\bm\delta})={\rm E}[{\bf D}^{-1}({\bm\gamma}-{\bm\delta})]$. It satisfies $\dot H_2({\bm 0})={\bm 0}$. The Hessian matrix of $H_2({\bm\delta})$ with respect to ${\bm\delta}$ is $\ddot H_2({\bm\delta})=-{\bf D}^{-1}$. It is negative definite. The gradient vector of $H({\bm\alpha},{\bm\delta})$ with respect to ${\bm\alpha}$ and ${\bm\delta}$ is $\dot H({\bm\alpha},{\bm\delta})=(\dot H_1^\top({\bm\alpha}),\dot H_2^\top({\bm\delta}))^\top$. It satisfies $\dot H({\bm 0},{\bm 0})={\bm 0}$. The Hessian matrix of $H({\bm\alpha},{\bm\delta})$ with respect to ${\bm\alpha}$ and ${\bm\delta}$ is $\ddot H({\bm\alpha},{\bm\delta})={\rm diag}(\ddot H({\bm\alpha}),\ddot H_2({\bm\delta}))$. It is negative definite. The local maximizer of $H({\bm\alpha},{\bm\delta})$ is unique almost surely, which can only be ${\bm\alpha}\stackrel{a.s}=0$ and ${\bm\delta}\stackrel{a.s}=0$. We conclude. \qed

\noindent
{\bf Proof of Theorem~\ref{thm:main theorem posterior mean and covariance}.} Let ${\bm v}_{\tilde{\bm\xi}}={\bf D}{\bf Z}^\top{\bf R}_{\tilde{\bm\xi}}^{-1}({\bm u}_{\tilde{\bm\xi}}-{\bf X}{\bm\beta}) $ and ${\bf V}_{\tilde{\bm\xi}}={\bf D}-{\bf D}{\bf Z}^\top{\bf R}_{\tilde{\bm\xi}}^{-1}{\bf Z}{\bf D}$ with ${\bm u}_{\tilde{\bm\xi}}={\bm\eta}_{\tilde{\bm\xi}}+[{\bm y}-b'({\bm\eta}_{\tilde{\bm\xi}})]/b''({\bm\eta}_{\tilde{\bm\xi}})$, ${\bf R}_{\tilde{\bm\xi}}={\bf Z}{\bf D}{\bf Z}^\top+{\bf W}_{\tilde{\bm\xi}}^{-1}$, ${\bf W}_{\tilde{\bm\xi}}={\rm diag}[b''({\bm\eta}_{\tilde{\bm\xi}})]$,  and ${\bm\eta}_{\tilde{\bm\xi}}={\bf X}{\bm\beta}+{\bf Z}{\tilde{\bm\xi}}$, where $\tilde{\bm\xi}$ can depend on ${\bm y}$, ${\bf X}$, and ${\bf Z}$ but not ${\bm\gamma}$. Based on the formulation of $\psi({\bm\delta})$, we consider the properties of
$$ \Psi({\bm\delta};\tilde{\bm\xi})=\int_{\mathbb{R}^n}\int_{\mathbb{R}^r} \log[\varphi({\bm\gamma}; {\bm v}_{\tilde{\bm\xi}}+{\bm\delta},{\bf V}_{\tilde{\bm\xi}})]  q({\bm\gamma}|{\bm y})\bar f({\bm y})d{\bm\gamma}\nu(d{\bm y}). $$
We compare the gradient vector of $\Psi({\bm\delta};\tilde{\bm\xi})$ with respect to ${\bm\delta}$ as
$$\dot\Psi_1({\bm\delta};\tilde{\bm\xi})=\int_{\mathbb{R}^n}\int_{\mathbb{R}^r} [{\bf V}_{\tilde{\bm\xi}}^{-1}({\bm\gamma}-{\bm\delta}-{\bm v}_{\tilde{\bm\xi}})]  q({\bm\gamma}|{\bm y})\bar f({\bm y})d{\bm\gamma}\nu(d{\bm y})$$
and the gradient vector of $\psi({\bm\delta})$ with respect to ${\bm\delta}$ as
$$\dot\psi({\bm\delta})=\int_{\mathbb{R}^n}\int_{\mathbb{R}^r} [{\bf V}^{-1}({\bm\gamma}-{\bm\delta}-{\bm v})]  q({\bm\gamma}|{\bm y})\bar f({\bm y})d{\bm\gamma}\nu(d{\bm y}).$$ 
The Hessian matrix of $\Psi({\bm\delta};\tilde{\bm\xi})$ with respect to ${\bm\delta}$ is negative definite. By $\dot\psi({\bm 0})={\bf 0}$ and the Bolzano–Poincaré–Miranda theorem~\cite[e.g.]{vrahatis2016}, there exists a unique spatial case of $\tilde{\bm\xi}$ such that $\Psi({\bm\delta};\tilde{\bm\xi})$ satisfies $\dot\Psi({\bm 0};\tilde{\bm\xi})={\bf 0}$, implying that ${\bm\delta}={\bf 0}$ is the unique maximizer of $\Psi({\bm\delta};\tilde{\bm\xi})$ in the special case of $\tilde{\bm\xi}$.  Because $\varphi({\bm\gamma}; {\bm v}_{\tilde{\bm\xi}}+{\bm\delta},{\bf V}_{\tilde{\bm\xi}})$ is a normal PDF, ${\bm v}_{\tilde{\bm\xi}}$ is the posterior mean and ${\bf V}_{\tilde{\bm\xi}}$ is the posterior covariance. The posterior mean is the mode of integrand of $\Psi({\bm 0};\tilde{\bm\xi})$. We consider 
$$\eqalign{
\tilde\Psi({\bm\alpha},{\bm\delta};\tilde{\bm\xi})=&\int_{\mathbb{R}^r}\int_{\mathbb{R}^n} \log[\varphi({\bm u}_{\tilde{\bm\xi}};{\bm\alpha}+{\bf X}{\bm\beta}+{\bf Z}{\bm\gamma},{\bf W}_{\tilde{\bm\xi}}^{-1})\varphi({\bm\gamma};{\bm\delta},{\bf D})]f({\bm y}|{\bm\gamma})\pi({\bm\gamma})\nu(d{\bm y})d{\bm\gamma}\cr
=&\int_{\mathbb{R}^r}\int_{\mathbb{R}^n} \log[\varphi({\bm u}_{\tilde{\bm\xi}};{\bm\alpha}+{\bf X}{\bm\beta}+{\bf Z}{\bm\gamma},{\bf W}_{\tilde{\bm\xi}}^{-1})]f({\bm y}|{\bm\gamma})\pi({\bm\gamma})\nu(d{\bm y})d{\bm\gamma}\cr
&+\int_{\mathbb{R}^r}\int_{\mathbb{R}^n} \log[\varphi({\bm\gamma};{\bm\delta},{\bf D})]f({\bm y}|{\bm\gamma})\pi({\bm\gamma})\nu(d{\bm y})d{\bm\gamma}\cr
}$$
for arbitrary ${\bm\alpha}$ and ${\bm\delta}$. It satisfies $\tilde\Psi(-{\bf Z}{\bm\delta},{\bm\delta};\tilde{\bm\xi})=\Psi({\bm\delta};\tilde{\bm\xi})$. Thus, $\tilde\Psi({\bm\alpha},{\bm\delta};\tilde{\bm\xi})\le\tilde\Psi({\bm 0},{\bm 0};\tilde{\bm\xi})=\Psi({\bm 0};\tilde{\bm\xi})$. The inequality holds if and only if ${\bm\alpha}\stackrel{a.s}={\bm 0}$ and ${\bm\delta}\stackrel{a.s.}={\bm 0}$.  We examine the first term of the above. We obtain
$$\eqalign{ 
&\log[\varphi({\bm u}_{\tilde{\bm\xi}};{\bm\alpha}+{\bf X}{\bm\beta}+{\bf Z}{\bm\gamma},{\bf W}_{\tilde{\bm\xi}}^{-1})]\cr
=&-{n\over 2}\log(2\pi)-{1\over 2}\log|{\bf W}_{\tilde{\bm\xi}}|-{1\over 2}[{{\bm y}-b'({\bm\eta}_{\tilde{\bm\xi}})\over b''({\tilde{\bm\xi}})}-{\bm\alpha}-{\bf Z}({\bm\gamma}-\tilde{\bm\xi})]^\top{\bf W}_{\tilde{\bm\xi}} [{{\bm y}-b'({\bm\eta}_{\tilde{\bm\xi}})\over b''({\tilde{\bm\xi}})}-{\bm\alpha}-{\bf Z}({\bm\gamma}-\tilde{\bm\xi})].
}$$
${\bm\alpha}={\bm 0}$ achieves the maximum under the integral. Based on the roles of ${\bf W}_{\tilde{\bm\xi}}$ in the normal likelihood, the maximum satisfies ${\rm E}({\bm\gamma}|{\bm y})=\tilde{\bm\xi}$. The special case induces $\tilde{\bm\xi}={\bm\xi}$, implying that ${\bm v}_{{\bm\xi}}$ is the posterior mean and ${\bf V}_{{\bm\xi}}$ is the posterior covariance. We conclude.  \qed

\noindent
{\bf Proof of Corollary~\ref{cor:posterior mean and covariance computation exact answers}}. Let ${\bm\delta}^{(t)}={\bm\xi}^{(t)}-{\bm\xi}^{(t-1)}$. The value of $\Psi({\bm\delta}^{(t)};{\bm\xi}^{(t)})$ increases in the iterations of the algorithm, implying that $\Psi({\bm\delta}^{(t)};{\bm\xi}^{(t)})$ approaches a local optimizer of $\Psi({\bm\delta},\tilde{\bm\xi})$. Note that $\Psi({\bm\delta},\tilde{\bm\xi})$ is uniquely maximized at ${\bm\delta}={\bm 0}$ and $\tilde{\bm\xi}={\bm\xi}$. We obtain ${\bm\delta}^{(t)}\rightarrow 0$ and $\tilde{\bm\xi}^{(t)}\rightarrow{\bm\xi}$. It induces ${\bm\Xi}_{{\bm\xi}^{(t)}}\rightarrow{\bm\Xi}$ in the iterations of the algorithm. We conclude. \qed

\end{document}